\documentstyle[12pt]{article}

\catcode`\@=11
\def\@citex[#1]#2{%
\if@filesw \immediate \write \@auxout {\string \citation {#2}}\fi
\@tempcntb\m@ne \let\@h@ld\relax \def\@citea{}%
\@cite{%
  \@for \@citeb:=#2\do {%
    \@ifundefined {b@\@citeb}%
      {\@h@ld\@citea\@tempcntb\m@ne{\bf ?}%
      \@warning {Citation `\@citeb ' on page \thepage \space undefined}}%
      {\@tempcnta\@tempcntb \advance\@tempcnta\@ne%
      \@tempcntb\number\csname b@\@citeb \endcsname \relax%
      \ifnum\@tempcnta=\@tempcntb 
        \ifx\@h@ld\relax%
          \edef \@h@ld{\@citea\csname b@\@citeb\endcsname}%
        \else%
          \edef\@h@ld{\ifmmode{-}\else--\fi\csname b@\@citeb\endcsname}%
        \fi%
      \else
        \@h@ld\@citea\csname b@\@citeb \endcsname%
        \let\@h@ld\relax%
      \fi}%
    \def\@citea{,\penalty\@highpenalty\,}%
  }\@h@ld
}{#1}}

\def\@citeb#1#2{{[#1]\if@tempswa , #2\fi}}
\def\@citeu#1#2{{$^{#1}$\if@tempswa , #2\fi }}
\def\@citep#1#2{{#1\if@tempswa , #2\fi}}

\def\bcites{         
        \catcode`\@=11
        \let\@cite=\@citeb
        \catcode`\@=12
}

\def\upcites{         
        \catcode`\@=11
        \let\@cite=\@citeu
        \catcode`\@=12
}

\def\plaincites{      
        \catcode`\@=11
        \let\@cite=\@citep
        \catcode`\@=12
}

\newcount\hour
\newcount\minute
\newtoks\amorpm
\hour=\time\divide\hour by 60
\minute=\time{\multiply\hour by 60 \global\advance\minute by-\hour}
\edef\standardtime{{\ifnum\hour<12 \global\amorpm={am}%
        \else\global\amorpm={pm}\advance\hour by-12 \fi
        \ifnum\hour=0 \hour=12 \fi
        \number\hour:\ifnum\minute<10 0\fi\number\minute\the\amorpm}}
\edef\militarytime{\number\hour:\ifnum\minute<10 0\fi\number\minute}

\def\draftlabel#1{{\@bsphack\if@filesw {\let\thepage\relax
   \xdef\@gtempa{\write\@auxout{\string
      \newlabel{#1}{{\@currentlabel}{\thepage}}}}}\@gtempa
   \if@nobreak \ifvmode\nobreak\fi\fi\fi\@esphack}
        \gdef\@eqnlabel{#1}}
\def\@eqnlabel{}
\def\@vacuum{}
\def\marginnote#1{}
\def\draftmarginnote#1{\marginpar{\raggedright\scriptsize\tt#1}}
\def\draft{
        \pagestyle{plain}
        \overfullrule=2pt
        \oddsidemargin -.5truein
        \def\@oddhead{\sl \phantom{\today\quad\militarytime} \hfil
        \smash{\Large\sl DRAFT} \hfil \today\quad\militarytime}
        \let\@evenhead\@oddhead
        \let\label=\draftlabel
        \let\marginnote=\draftmarginnote
        \def\ps@empty{\let\@mkboth\@gobbletwo
        \def\@oddfoot{\hfil \smash{\Large\sl DRAFT} \hfil}
        \let\@evenfoot\@oddhead}
        \def\@eqnnum{(\theequation)\rlap{\kern\marginparsep\tt\@eqnlabel}%
        \global\let\@eqnlabel\@vacuum}  }

\def\blackfonts{
        \font\blackboard=msbm10 scaled\magstep1
        \font\blackboards=msbm8
        \font\blackboardss=msbm6
}

\def\prep{         
        \catcode`\@=11
        \input art10.sty
        \catcode`\@=12
        
        \let\small\null
        \def\blackfonts{
                \font\blackboard=msbm10
                \font\blackboards=msbm7
                \font\blackboardss=msbm5
        }
        \let\sl\it
        \twocolumn
        \sloppy
        \voffset=-2.54truecm
        \hoffset=-2.54truecm
        \flushbottom
        \parindent 1em
        \leftmargini 2em
        \leftmarginv .5em
        \leftmarginvi .5em
        \marginparwidth 48pt
        \marginparsep 10pt
        \setlength{\columnsep}{2truecm}
        \setlength{\textwidth}{25.4truecm}
        \setlength{\textheight}{17truecm}
        \baselineskip=16pt
        \oddsidemargin .18truein
        \evensidemargin .17truein
}

\def\eqalign#1{\null\,\vcenter{\openup\jot\m@th
  \ialign{\strut\hfil$\displaystyle{##}$&$\displaystyle{{}##}$\hfil
      \crcr#1\crcr}}\,}
\def\eqalignno#1{\displ@y \tabskip\centering
  \halign to\displaywidth{\hfil$\@lign\displaystyle{##}$\tabskip\z@skip
    &$\@lign\displaystyle{{}##}$\hfil\tabskip\centering
    &\llap{$\@lign##$}\tabskip\z@skip\crcr
    #1\crcr}}

\def\section{\@startsection {section}{1}{\z@}{3.ex plus 1ex minus
 .2ex}{2.ex plus .2ex}{\large\bf}}
\def\subsection{\@startsection{subsection}{2}{\z@}{2.75ex plus 1ex minus
 .2ex}{1.5ex plus .2ex}{\bf}}        

\def\appendix{{\newpage\section*{Appendix}}\let\appendix\section%
        {\setcounter{section}{0}
        \gdef\thesection{\Alph{section}}}\section}

\def\abstract{\if@twocolumn
\section*{Abstract}
\else 
\begin{center}
{\bf Abstract\vspace{-.5em}\vspace{0pt}}
\end{center}
\quotation
\fi}

\catcode`\@=12

\def\d{\partial}

\def\sqr#1#2{{\vcenter{\vbox{\hrule height.#2pt\hbox{\vrule width.#2pt 
height#1pt \kern#1pt \vrule width.#2pt}\hrule height.#2pt}}}}
\def\w{\mathchoice\sqr45\sqr45\sqr{2.1}3\sqr{1.5}3\,}

\def\=d{\,{\buildrel\rm def\over =}\,}

\def\F{{\cal F}}

\def\S{\hbox{\bbf S}}

\def\i3p{\p32\int d^3p}

\def\As{A\hbox to 1pt{\hss /}}
\def\np4{\int d^4p_1\cdots d^4p_{n-1}\, }

\def\nx4{\int d^4x_1\ldots d^4x_n\, }

\def\kon#1#2{\vbox{\halign{##&&##\cr
\lower4pt\hbox{$\scriptscriptstyle\vert$}\hrulefill &
\hrulefill\lower4pt\hbox{$\scriptscriptstyle\vert$}\cr $#1$&
$#2$\cr}}}

\def\konv#1#2#3{\hbox{\vrule height12pt depth-1pt}
\vbox{\hrule height12pt width#1cm depth-11.6pt}
\hbox{\vrule height6.5pt depth-0.5pt}
\vbox{\hrule height11pt width#2cm depth-10.6pt\kern5pt
      \hrule height6.5pt width#2cm depth-6.1pt}
\hbox{\vrule height12pt depth-1pt}
\vbox{\hrule height6.5pt width#3cm depth-6.1pt}
\hbox{\vrule height6.5pt depth-0.5pt}}
\def\konu#1#2#3{\hbox{\vrule height12pt depth-1pt}
\vbox{\hrule height1pt width#1cm depth-0.6pt}
\hbox{\vrule height12pt depth-6.5pt}
\vbox{\hrule height6pt width#2cm depth-5.6pt\kern5pt
      \hrule height1pt width#2cm depth-0.6pt}
\hbox{\vrule height12pt depth-6.5pt}
\vbox{\hrule height1pt width#3cm depth-0.6pt}
\hbox{\vrule height12pt depth-1pt}}

\def\konw#1#2#3{\hbox{\vrule height12pt depth-1pt}
\vbox{\hrule height12pt width#1cm depth-11.6pt}
\hbox{\vrule height6.5pt depth-0.5pt}
\vbox{\hrule height12pt width#2cm depth-11.6pt \kern5pt
      \hrule height6.5pt width#2cm depth-6.1pt}
\hbox{\vrule height6.5pt depth-0.5pt}
\vbox{\hrule height12pt width#3cm depth-11.6pt}
\hbox{\vrule height12pt depth-1pt}}

\def\i{{\rm int}}

\def\e{{\rm ext}}

\def\r{{\rm ret}}
\def\a{{\rm av}}

\def\m3{{\mu_1\mu_2\mu_3}}

\def\p{{(+)}}

\def\be{\begin{equation}}       \def\eq{\begin{equation}}
\def\ee{\end{equation}}         \def\eqe{\end{equation}}

\def\bea{\begin{eqnarray}}      \def\eqa{\begin{eqnarray}}
\def\ena{\end{eqnarray}}        \def\eea{\end{eqnarray}}
                                \def\eqae{\end{eqnarray}}

\def\ba{\begin{array}}
\def\ea{\end{array}}
\def\unit{1 \hskip-.3em \raise2pt\hbox{$ \scriptstyle |$ } }

\def\a{\alpha}
\def\b{\beta}
 
\def\d{\delta}
\def\e{\epsilon}           
\def\f{\phi}               

\def\i{\iota}

\def\k{\kappa}                    
\def\l{\lambda}
\def\m{\mu}
\def\n{\nu}
\def\o{\omega}  \def\w{\omega}
\def\p{\pi}                
\def\r{\rho}                                     
\def\s{\sigma}                                   
\def\t{\tau}

\def\F{\Phi}
\def\G{\Gamma}

\def\L{\Lambda}
\def\O{\Omega}

\def\S{\Sigma}

\def\cl{{\cal L}}
\def\cm{{\cal M}}
\def\cn{{\cal N}}

\def\cq{{\cal Q}}

\def\half{{1 \over 2}}

\def\bop#1{\setbox0=\hbox{$#1M$}\mkern1.5mu
        \vbox{\hrule height0pt depth.04\ht0
        \hbox{\vrule width.04\ht0 height.9\ht0 \kern.9\ht0
        \vrule width.04\ht0}\hrule height.04\ht0}\mkern1.5mu}
\def\Box{{\mathpalette\bop{}}}                        
\def\pa{\partial}                              

\def\>{\rangle} 

\def\<{\langle} 
\def\Dsl{D \hskip-.6em \raise1pt\hbox{$ / $ } }

\def\sl#1{\rlap{\hbox{$\mskip 1 mu /$}}#1}
\def\leftrightarrowfill{$\mathsurround=0pt \mathord\leftarrow \mkern-6mu
       \cleaders\hbox{$\mkern-2mu \mathord- \mkern-2mu$}\hfill
       \mkern-6mu \mathord\rightarrow$}
\def\dvec#1{\vbox{\ialign{##\crcr
       \leftrightarrowfill\crcr\noalign{\kern-1pt\nointerlineskip}
       $\hfil\displaystyle{#1}\hfil$\crcr}}}          
\def\hook#1{{\vrule height#1pt width0.4pt depth0pt}}
\def\leftrighthookfill#1{$\mathsurround=0pt \mathord\hook#1
       \hrulefill\mathord\hook#1$}
\def\underhook#1{\vtop{\ialign{##\crcr                 
       $\hfil\displaystyle{#1}\hfil$\crcr
       \noalign{\kern-1pt\nointerlineskip\vskip2pt}
       \leftrighthookfill5\crcr}}}
\def\smallunderhook#1{\vtop{\ialign{##\crcr      
       $\hfil\scriptstyle{#1}\hfil$\crcr
       \noalign{\kern-1pt\nointerlineskip\vskip2pt}
       \leftrighthookfill3\crcr}}}

\def\sfrac#1#2{{\vphantom1\smash{\lower.5ex\hbox{\small$#1$}}\over
       \vphantom1\smash{\raise.4ex\hbox{\small$#2$}}}} 
\def\bfrac#1#2{{\vphantom1\smash{\lower.5ex\hbox{$#1$}}\over
       \vphantom1\smash{\raise.3ex\hbox{$#2$}}}}      
\def\afrac#1#2{{\vphantom1\smash{\lower.5ex\hbox{$#1$}}\over#2}}  
\def\on#1#2{{\buildrel{\mkern2.5mu#1\mkern-2.5mu}\over{#2}}}
\def\ddt#1{\on{\hbox{\LARGE .\kern-2pt.}}#1}             
\def\tdt#1{\on{\hbox{\LARGE .\kern-2pt.\kern-2pt.}}#1}   

\def\boxes#1{
       \newcount\num
       \num=1
       \newdimen\downsy
       \downsy=-1.5ex
       \mskip-2.8mu
       \bo
       \loop
       \ifnum\num<#1
       \llap{\raise\num\downsy\hbox{$\bo$}}
       \advance\num by1
       \repeat}
\def\boxup#1#2{\newcount\numup
       \numup=#1
       \advance\numup by-1
       \newdimen\upsy
       \upsy=.75ex
       \mskip2.8mu
       \raise\numup\upsy\hbox{$#2$}}

\newskip\humongous \humongous=0pt plus 1000pt minus 1000pt
\def\caja{\mathsurround=0pt}
\def\eqalign#1{\,\vcenter{\openup2\jot \caja
       \ialign{\strut \hfil$\displaystyle{##}$&$
       \displaystyle{{}##}$\hfil\crcr#1\crcr}}\,}
\newif\ifdtup

\def\PRD{Phys. Rev. D}

\def\PRL#1#2#3{Phys. Rev. Lett. {\bf#1} (#2) #3}

\def\NPB#1#2#3{Nucl. Phys. {\bf B#1} (#2) #3}

\def\PRD#1#2#3{ Phys. Rev. {\bf D#1} (#2) #3}

\def\PLB#1#2#3{Phys. Lett. {\bf #1B} (#2) #3}

\def\to{\rightarrow}

\def\1ov4{{1\over 4}}

\def\pa{\partial}

\def\ddt{\dot{\t}}

\def\pa{\partial}

\def\bb{\bar{b}}

\def\nonu{\nonumber \\{}}
\def\half{{1 \over 2}}

\font\mybb=msbm10 at 12pt
\def\bb#1{\hbox{\mybb#1}}

\def\bE {\bb{E}}

\def\IR{\relax{\rm  I\kern-.18em R}}

\def\ind{{1\over 4 \pi \a'}\int\!\! d^2 \s}
\def\swa{$\swarrow$}
\def\dar{$\downarrow$}

\begin{document}

\thispagestyle{empty} \setcounter{footnote}{0}
\begin{flushright}
SPIN-1998/17\\ 
hep-th/9901050
\end{flushright}
\vskip 1,7cm
\begin{center}
\begin{center}
{\large\bf Black holes and branes in string theory}
\end{center}
\vskip 2.0cm
  {\bf Kostas Skenderis} 
\vskip 0.5cm
  {\it Spinoza Institute, University of Utrecht,\\ 
 Leuvenlaan 4, 3584 CE Utrecht, The Netherlands \\
\vskip 0.2cm
{\tt K.Skenderis@phys.uu.nl}}
\end{center}
\vskip 1,0cm
{\bf Abstract.} -
This is a set of introductory lecture notes on black holes in 
string theory. After reviewing some aspects of string theory 
such as dualities, brane solutions, 
supersymmetric and non-extremal intersection rules,  
we analyze in detail extremal and non-extremal $5d$ black holes.
We first present the D-brane counting for extremal black holes.
Then we show that $4d$ and $5d$ non-extremal black holes can be 
mapped to the BTZ black hole (times a compact manifold) by means of dualities.
The validity of these dualities is analyzed in detail. 
We present an analysis
of the same system in the spirit of the adS/CFT correspondence.
In the ``near-horizon'' limit (which is actually a near inner-horizon limit
for non-extremal black holes) the black hole reduces again to the 
BTZ black hole. A state counting is presented in terms of the BTZ 
black hole.

\newpage

\section{Introduction}

The physics of 20th century is founded on two pillars: 
quantum theory and general theory of relativity. 
Quantum theory has been extremely successful in describing the
physics at microscopic scales while general relativity has been 
equally successful with physics at cosmological scales. 
However, attempts to construct a quantum theory of gravity 
stubble upon the problem of the non-renormalizability of the theory. 
Is it really necessary to have a quantum theory of gravity? Why not 
having gravity classical and matter quantized? Is it just an aesthetic 
question or is there an internal inconsistency if some of the physical laws 
are classical and some quantum? If some of the interactions are classical 
then one could use only these interactions in order to arbitrarily 
obtain the position and the velocity of particles, thus violating Heiseberg's
uncertainty principle.  Therefore, at the fundamental level, 
if some of the physical laws are quantum, all of them have to be quantum.

It is amusing to see what happens if we insist on both classical 
general relativity and the uncertainty principle. Suppose we want 
to measure a spacetime coordinate with accuracy $\d x$, then 
by the uncertainty principle there will be energy of order 
$1/\d x$ localized in this region. But if $\d x$ is very small 
then the energy will be so large that a black hole will be 
formed, and the spacetime point will be hidden behind a horizon!
One can estimate\cite{DFR} that the scale that leads to a black hole 
formation (through the uncertainty principle) is of order of the Planck 
length $l_p$. Therefore, classical general relativity 
and quantum mechanics become incompatible at scales of order $l_p$.

One of the most fascinating objects that general relativity predicts
is black holes. Classically, black holes are completely black. Objects 
inside their event horizon are eternally trapped. Even light rays 
are confined by the gravitational force. In addition, there is 
a singularity hidden behind the horizon.
In the early seventies, a number of laws that govern the physics 
of black holes were established\cite{chris,Hawk2,Beke}. 
In particular, it was found that there is a very close analogy between 
these laws and the four laws of thermodynamics\cite{Hawk2}.
The black hole laws become that of thermodynamics
if one replaces the surface gravity $\k$ of the black hole by the 
temperature $T$ of a body in thermal equilibrium, 
the area of the black hole $A$ by the entropy $S$\cite{Beke}, 
the mass of the black hole $M$ by the energy of the system $E$ etc. 
It is natural to wonder whether this formal similarity is more than just 
an analogy. At the classical level one immediately runs into a problem 
if one tries to take this analogy seriously: classically black holes only 
absorb so their temperature is strictly zero. In a seminal paper\cite{Hawk}, 
however, Hawking showed that quantum 
mechanically black holes emit particles with thermal spectrum. The temperature 
was found to be $T=\k/2 \p$! Then from the first law follows
the ``Bekenstein-Hawking entropy formula'',
\be \label{BH}
S={A \over 4 G_N}
\ee
where $G_N$ is Newton's constant. 
Having established that black hole laws are thermodynamic in nature 
one would like to understand what is the underlying microscopic theory.
What are the microscopic degrees of freedom that make up the black hole?

Since black holes radiate, they lose mass and they may eventually 
evaporate. Observing such a phenomenon is rather unlikely 
since one can estimate the lifetime of a black hole of stellar mass 
to have lifetime\footnote{
For black holes of mass $M$ the Hawking temperature is of order 
$T \sim 10^{-6} (M_\odot /M) \ K$ and their lifetime 
of order $10^{71} (M_\odot /M)^{-3} \ s$.}
longer than the age of the universe.
The fact, however, that black holes Hawking radiate and may eventually 
evaporate leads to an important paradox. The matter that falls into 
black hole has structure. The outgoing radiation, however, is 
structure-less since it is thermal. What happens to the information 
stored in the black hole if the black hole completely evaporates?
If it gets lost then the evolution is not unitary. 
Hawking argued that these considerations imply that quantum mechanics 
has to be modified. There is great controversy over the question of the 
final state of black holes, and there is no completely satisfactory scenario. 
We will not enter into this question in this lectures. 
Let us note, however, that the 
resolution of this problem may be related to the question 
of understanding the microscopic description of black holes.
Radiation from stars also has a thermal spectrum. However, we do not claim
that information is lost in stars. The thermal spectrum is due to averaging 
over microscopic states. 

We have seen that semi-classical considerations yield a number of important
issues. Any consistent quantum theory of gravity should provide answers
to the questions raised in the previous paragraphs. The leading candidate for 
a quantum theory of gravity is string theory. Therefore, string theory ought 
to resolve these issues. Issues involving black holes are non-perturbative 
in nature. Up until recently, however, we only had a perturbative 
formulation of string theory. The situation changed dramatically over 
the last few years. Dualities have led to a unified
picture of all string theories \cite{HuTo,Wit1}. Moreover, new
non-perturbative objects, the D-branes, were discovered\cite{JPol}.
These new ingredients made possible to tackle some of the problems mentioned
above.

In this lectures we review recent progress in understanding 
black holes using string theory.
We start by briefly reviewing perturbative strings, D-branes and dualities
in section 2. In particular,  we review in some detail T-duality in 
backgrounds with isometries. 
In section 3 we present the brane solutions of type II 
and eleven dimensional supergravity, their connections through 
dualities, and a set of intersection rules that yields new solutions
describing configurations of intersecting branes. 
We use these results in section 4 in order 
to study extremal and non-extremal black 
holes. In section 4.1 we analyze extremal $5d$ black holes. We show that one
can derive the Bekenstein-Hawking entropy formula by counting 
D-brane states. In section 4.2 we show that $4d$ and $5d$ 
non-extremal black holes can be mapped to the BTZ black hole\cite{BTZ,BHTZ} 
(times a compact manifold) by means of dualities. 
We show that a general U-duality transformation
preserves the thermodynamic characteristics of black holes. Then we 
critically examine the so-called shift transformation that removes 
the constant part from harmonic functions. We show that this transformation 
also preserves the thermodynamic characteristics of the original 
black hole. In general, however, it is not a symmetry of the theory. 
Section 4.3 contains a short introduction to adS/CFT duality,
and its application to black holes. The low-energy decoupling limit 
employed in the adS/CFT correspondence (which is a near inner-horizon limit 
for non-extremal black holes) also yields a connection with the BTZ black hole.
We use the connection to the BTZ black hole to infer a state counting
for the higher dimensional black holes. 

Previous reviews for black holes 
in string theory include \cite{Hor1,malda1,peet}. 

\newpage
\section{String theory and dualities.}

In this section we present some aspects of string theory. 
The main purpose is to set our conventions and to review 
certain material that they will be of use in later sections.

\subsection{Bosonic string and D-branes}

The worldsheet action for the bosonic string is given by
\be
S={1 \over 4 \pi \a'} \int d \tau \int_0^\p d \s \sqrt{h}
h^{ab} \pa_a X^\m \pa_b X_\m.
\ee
where $h$ is the worldsheet metric. 
The tension of the string is given by $T=1/(2 \p \a')$
($\a'$ is the square of the string length $l_s$).
Varying the action we obtain 
\be \label{var}
\d S = - {1 \over 2 \p \a'} \int d \t d \s \sqrt{h} \d X^\m \Box X^\m  
+{1 \over 2 \p \a'} \int d \t [\sqrt{h} \pa_\s X_\m \d X^\m]_{\s=0}^{\s=\p}
\ee
In order to have a well-defined variational problem the last term should 
vanish. This implies three different types of boundary conditions
\bea
X^\m(\t, \s) = X^\m(\t, \s + \p) \qquad && \mbox{closed\ string} \nonu
\pa_\s X^\m(\s=0)=\pa_\s X^\m(\s=\p)=0 \qquad &&
\mbox{open\ string\ with\ Neumann\ BC}  \nonu
X^\m(\s=0)=const,\ X^\m(\s=\p)=const , \qquad&&
\mbox{open\ string\ with\ Dirichlet\ BC} \nonumber
\eea

The Neumann boundary conditions for the open string imply that there is no 
momentum flow at the end of the string. With Dirichlet boundary conditions,
however, there is momentum flowing from the string to the hypersurface 
where the string ends. Therefore, this hypersurface, the D-brane,
is a dynamical object. 

One may (first) quantize the string using standard methods. The closed
string consist of left and right movers. We denote the left and 
right level by $N$ and $\tilde{N}$, respectively. For open strings we have 
only one kind of oscillators.
The perturbative spectrum for the three kind of boundary conditions 
listed above is given by 
\bea \label{spec}
M^2_{closed}={2 \over \a'}(N+\tilde{N} -2) \nonu
M_{open, N}^2 = {1 \over \a'} (N -1) \nonu
M_{open, D}^2 = \left({l \over 2 \p \a'}\right)^2 + {1 \over \a'} (N -1)
\eea
The term $l/2 \p \a'{=}l T$ is the energy of a string of length $l$ 
stretched between two D-branes. 

From (\ref{spec}) follows that the massless spectrum of closed 
strings consist of a graviton $G_{\m \n}$, an antisymmetric tensor
$B_{\m \n}$ and a dilaton $\f$. The massless spectrum 
of open strings with Neumann boundary 
conditions consists of a photon $A_\m$. Finally, for a string that 
ends on a D$p$-brane, i.e. the open string endpoints are confined 
to the $p+1$-dimensional worldvolume of the D-brane, we get a vector
field $A_m$, $m=0, \ldots, p$,
that lives on the worldvolume of the D-brane, and $(25-p)$ scalars.
The latter encode the fluctuations of the position of the D-brane. 

The string coupling constant is not a new parameter but the expectation 
value of the dilaton field, $\< e^\f \> = g_s$. 
String theory perturbation theory is weighted by $g_s^\chi$, where $\chi$ 
is the Euler number of the
string worldsheet. A compact surface can be built 
by adding $g$ handles, $c$ cross-caps and $b$ boundaries to the 
sphere. Its Euler number is given by $\chi=2 - 2g -b -c$.  
Hence, the closed string coupling constant is proportional to
the square of the open string coupling constant.

One may calculate the tension of D-branes\cite{JPol,TASI}
\be
T_p \sim {1 \over g_s l_s^{p+1}}.
\ee
Since the tension of the D-brane depends on the inverse of the string 
coupling constant, D-branes are non-perturbative objects. Notice that
this behavior is different from the behavior of field
theory solitons whose mass goes as $1/g^2$, where $g$ is the field
theory coupling constant. The existence of such non-perturbative objects 
is required by string duality \cite{Wit1}.

\subsection{Superstrings}

There are five consistent string theories; type IIA and IIB, type I, heterotic
$SO(32)$ and heterotic $E_8 \times E_8$. All of them are related 
through dualities. In this review we shall concentrate on type II 
theories, so we briefly present some aspects of them.

The bosonic massless sector of type II theories consist of the following 
fields

\begin{tabular}{ccc}
Type IIA & \qquad $g_{\m \n} \ \ B_{\m \n} \ \ \f$ & \qquad
$C_\m^{(1)} \ \ C^{(3)}_{\m \n \l}$ \\
Type IIB & \qquad $g_{\m \n} \ \ B_{\m \n} \ \ \f$ & \qquad
$C^{(0)} \ \ C_{\m \n}^{(2)} \ \ C^{(4)\ +}_{\k \l \m \n}$, 
\end{tabular} \newline
where $C^{(p)}$ are $p$-index antisymmetric gauge fields. The $+$ 
in $C^{(4)+}$ indicates that the field strength is self-dual.
The graviton $g_{\m \n}$, the antisymmetric tensor $B_{\m \n}$ 
and the dilaton $\f$ make up the NSNS sector. These fields couple to 
perturbative strings. The RR sector (i.e. the
antisymmetric tensors $C^{(p+1)}$), however, does not couple to perturbative 
strings but rather to D$p$-branes.

Extended objects naturally couple to antisymmetric tensors.
The prototype example is the coupling of the point particle to electromagnetic
field, $\int A_\m dx^\m$. Similarly, fundamental string naturally couple
to $B_{\m \n}$, and D$p$-branes to $C^{(p+1)}$ 
\bea
&&\int_\S B_{\m \n} dx^\m \wedge dx^\n \nonu
&&\int_{\cm_{p+1}} C^{(p+1)}_{\m_1 \cdots \m_{p+1}} dx^{\m_1} \wedge 
\cdots \wedge dx^{\m_{p+1}}
\eea
where $\S$ and $\cm_{p+1}$ is the string worldsheet and D$p$-brane 
worldvolume, respectively. To each ``electric'' $p$-brane there 
is also a dual ``magnetic'' $(6-p)$-brane. 
(To see this notice that $*d C^{(p+1)} = d \tilde{C}^{(7-p)}$).
In particular, there is a solitonic 5-brane (NS5) that is the magnetic
dual of a fundamental string F1. 
In addition, strings can carry momentum. This corresponds in 
low energy to gravitational waves (W). The (Hodge) dual to 
waves are Kaluza-Klein monopoles (KK) (see section \ref{low}).  

In summary, we have the following objects in type II theory 
(D(-1) are D-instanton and D9 are spacetime-filling branes)

\begin{tabular}{ccc}
Type IIA & \qquad W\ \ F1\ \ NS5\ \ KK & \qquad
D0\ \ D2\ \ D4\ \ D6\ \ D8 \\
Type IIB & \qquad W\ \ F1\ \ NS5\ \ KK & \qquad
D(-1)\ \ D1\ \ D3\ \ D5\ \ D7\ \ D9
\end{tabular} \newline

We have deduced the existence of dynamical extended objects
by considering perturbative string theory. 
These states, however, preserve half of maximal supersymmetry  
and therefore continue to exit 
at all values of the string coupling constant. 

\subsection{Dualities}

A central element in the recent developments are the duality symmetries 
of string theory. The duality symmetries are believed to be exact 
discrete gauge symmetries spontaneously broken by scalar vev's.

The best-understood duality symmetry is T-duality. This symmetry is 
visible in string perturbation theory but it is non-perturbative 
on the worldsheet. T-duality relates compactifications on a manifold
of (large) volume $v$ to compactifications
on a manifold of (small) volume $1/v$.
The simplest case is compactification on a circle. Upon such compactification 
the two type II theories, and heterotic $E_8 \times E_8$ and heterotic
$SO(32)$ theories are equivalent,
\bea
[IIA]_{R} &\stackrel{T}{\longleftrightarrow}& [IIB]_{1/R} \nonu
[Het\ E_8 \times E_8]_R  &\stackrel{T}{\longleftrightarrow}& 
[Het\ SO(32)]_{1/R}, \nonumber
\eea
where the subscript indicates that the theory is compactified on 
a circle of radius $R$ ($1/R$).

The action of T-duality on the various objects present in II theories 
is given in Table 1. The T-duality may be performed along one
of the worldvolume directions or along a transverse direction (for 
the KK monopole the transverse direction is taken to be the nut direction (see 
section \ref{low})).  
\begin{table}[h]
\begin{center}
\vspace{.2cm}
\begin{tabular}{|c|c|c|}
\hline
          & Parallel    &  transverse        \\ \hline\hline
D$p$      & D$(p-1)$    &  D$(p+1)$          \\ \hline
F1        & W           &  F1                \\ \hline
W         & F1          &  W                 \\ \hline
NS5       & NS5         &  KK                \\ \hline
KK        & KK          &  NS5               \\ \hline
\end{tabular}
\caption{{\it T-duality along parallel and transverse directions}}
\end{center}
\end{table}
More generally, T-duality 
asserts that different spacetimes with isometries may be equivalent
in string theory. We shall present the argument in some detail 
in the next section since we will make use of these results. 

A (conjectured) non-perturbative symmetry is S-duality. This is 
non-perturbative because it acts on the dilaton as $g_s \to 1/g_s$.
Thus, S-duality relates the strong coupling regime of one theory 
to the weak coupling regime of another. In particular we have
\bea
IIB &\stackrel{S}{\longleftrightarrow}& IIB \nonu
Het\ SO(32) &\stackrel{S}{\longleftrightarrow}& Type\ I
\eea
Actually, IIB string theory is believed to have an exact non-perturbative
$SL(2,Z)$ symmetry. In the following we shall only make use of the 
$Z_2$ subgroup that sends $\tau=C^{(0)} + i e^{-\phi}$ to $-1/\tau$,
interchanges $B_{\m \n}$ with $C^{(2)}_{\m \n}$, and leaves invariant 
$C^{(4)+}$ (so, in terms of branes, S-duality interchanges F1 with D1,
NS5 with D5, and leaves invariant the D3 brane).

S-duality allows one to get a handle to the strong 
coupling limit of three of the five string theories. 
In turns out that the strong coupling limit of IIA and heterotic 
$E_8 \times E_8$ theories is of a more ``exotic'' nature. One
gets instead an 11 dimensional theory, the M-theory\cite{Wit1,To}. 
M-theory on a small circle of radius $R_{11} = g_s l_s$ yields
IIA theory with string coupling constant $g_s$\cite{Wit1}.
Since perturbative string theory is an expansion around $g_s=0$,
the eleventh dimensions is not visible perturbatively.
Likewise, M-theory on an interval gives 
$E_8 \times E_8$ string theory\cite{HoWi1}.
Actually, all string theories can be obtained in suitable limits
from eleven dimensions. 

Although we do not have a fundamental understanding of what M-theory
is, we know that in low-energies M-theory reduces to 11 dimensional 
supergravity\cite{11sugra}. Eleven-dimensional supergravity compactified
on a torus yields a lower dimensional Poincar\'{e} supergravity with a 
certain duality group. The discretized version of this duality 
group is conjectured\cite{HuTo} (and widely believed) to be an 
exact symmetry of M-theory. T and S duality combine to yield
this bigger group, the U-duality group.
  
\subsubsection{Buscher's duality} \label{bsec}

Consider the sigma model 
\be
S=\ind \, \sqrt{h} 
[(h^{ab} g_{\m \n}+ i {\e^{ab} \over \sqrt{h}}
B_{\m \n})\pa_a X^\m \pa_b X^\n 
+ \a' R^{(2)} \phi] \, ,
\label{ssigma}
\ee
where $h$ and $R^{(2)}$ is the worldsheet metric and curvature, 
$g$ is the target space metric and $B$ is a potential
for the torsion 3-form $H=dB$. 
This action is invariant under the
transformation
\begin{equation}
\label{isometry}
\delta X^\m=\epsilon k^\m
\end{equation}
when the vector field $k^\m$ is a Killing vector, the Lie derivative of 
$B$ is a total derivative and the dilaton is invariant,
\bea
&&\cl_k g_{ij}=k_{i;j}+k_{j;i}=0, \nonu
&&\cl_k B =  \i_k dB +d \i_k B = d(v + \i_k B) \nonu
&&\cl_k \phi = k^\m \pa_\m \phi=0
\eea
One can now choose adapted coordinates 
$\{X^\m\}=\{ x,x^i \}$ such that the isometry acts
by translation of $x$, and all fields $g$, $B$ and $\phi$
are independent of $x$. In adapted coordinates, 
the killing vector is equal to 
$k^\m \pa/ \pa X^\m = \pa / \pa x$.  

To obtain the dual theory we first gauge the symmetry and add a Lagrange 
multiplier $\chi$ that imposes that the gauge connection is flat
\cite{Bus1}.
The result (in the conformal gauge and omitting the dilaton term)
is (see \cite{RV}, \cite{AABL} for details)
\be \label{sgauged}
S_1={1 \over 2 \p \a'}\int d^2z 
[(g_{\m \n} + B_{\m \n}) \pa X^\m \bar{\pa} X^\n
+(J_k - \pa \chi) \bar{A} + (\bar{J}_k + \bar{\pa} \chi) A + k^2 A \bar{A}]
\ee
where $J_k=(k+v)_\m \pa X^\m$, $\bar{J}_k=(k-v)_\m \bar{\pa} X^\m$ are the 
components of the Noether current associated with the symmetry.
If one integrates out the Lagrange multiplier field $\chi$, on a
topologically trivial worldsheet the gauge fields are pure gauge, 
$A = \pa \theta$, $\bar{A} = \bar{\pa} \theta$, and
one recovers the original model (\ref{ssigma}).  

If one integrates out the gauge fields $A , \bar{A}$ one finds the {\it dual\/}
model. One obtains (\ref{ssigma}) but with dual
background fields $\tilde{g}, \tilde{B}, \tilde{\F}$.
In adapted coordinates $\{X^\m\}=\{x, x^i\}$, 
\bea
\label{dual1}
&&\tilde{g}_{xx} = \frac{1}{g_{xx}} \qquad 
\tilde{g}_{xi} = \frac{B_{xi}}{g_{xx}} \qquad
\tilde{g}_{ij} = g_{ij} - \frac{g_{xi}g_{xj}-B_{xi}B_{xj}}{g_{xx}} 
\nonu
&&\tilde{B}_{xi} = \frac{g_{xi}}{g_{xx}} \qquad
\tilde{B}_{ij} = B_{ij} + \frac{g_{xi}B_{xj}-B_{xi}g_{xj}}{g_{xx}} \nonu 
&&\tilde{\phi} = \phi -\half \ln g_{xx}
\eea
The dilaton shift is a quantum mechanical effect \cite{Bus2} (see 
\cite{DRST} for recent careful discussion).

Another useful way to write these transformation rules is to 
re-write the metric as   
\be \label{KKmetric}
ds^2 = g_{xx}(dx + A_i dx^i)^2 + \bar{g}_{ij} dx^i dx^j
\ee
where $A_i = g_{xi}/g_{xx}$. Then the duality transformations take the 
form\cite{hor1}
\bea \label{dual2}
&&\tilde{g}_{xx} = \frac{1}{g_{xx}}, \qquad \tilde{A}_i = B_{xi}, \qquad
\tilde{B}_{xi}=A_i, \qquad \tilde{B}_{ij}=B_{ij}-2A_{[i} B_{j]x} \nonu
&&\tilde{\phi}=\phi - \half \ln g_{xx} \qquad \bar{g}_{ij} \ \mbox{invariant}
\eea
This form of the transformation rules exhibits most clearly the spacetime 
interpretation of the duality transformations. The form of the metric in 
(\ref{KKmetric}) is the standard KK ansatz for reduction over $x$.
Dimensional reduction over $x$  leads to a
$(d{-}1)$-dimensional theory which is 
invariant under the transformations in (\ref{dual2}).
These transformations act only on the matter fields 
and not on the pure gravitational sector.

Let us now discuss under which conditions the dual models are truly 
equivalent as conformal field theories.  

$\bullet$ {\it Compact vs non-compact isometries} \newline
In our discussion above we assumed that the worldsheet is trivial.
Let us relax this condition. Suppose also that we deal 
with a compact isometry.
The constraint on $A ,\bar{A}$ that comes from integrating out
the Lagrange multiplier $\chi$ implies $A,\bar{A}$ are flat, but in
principle they still may have nontrivial holonomies around
non-contractible loops. These holonomies can be constrained to 
vanish if $\chi$ has appropriate period\cite{RV,AABL}.
In summary, dualizing along a compact isometry one obtains a dual 
geometry which also has a compact isometry. The 
periods of the original and dual coordinate are reciprocal to each 
other. If this condition does not hold,  the
two models are not fully equivalent but related via an orbifold
construction. 

Non-compact isometries can be considered as a
limiting case. Since in this case $x$ takes any real value,
the dual coordinate $\chi$ must have period zero.
The dual manifold is an orbifold obtained by modding
out the translations in $\chi$. 
 
$\bullet$ {\it Isometries with fixed points} \newline
In our analysis we also assumed that the isometry is spacelike.
If the isometry is timelike then it follows from (\ref{sgauged})
that the integration over the gauge field yields a divergent
factor. If the isometry is null then the quadratic in the 
gauge field term in (\ref{sgauged}) vanishes. Therefore
these cases require special attention.
We refer to \cite{Tred,CLLPST,JN} for work concerning 
dualization (or the closely related issue of dimensional 
reduction) along timelike or null isometries.

A spacelike isometry may act freely or have fixed points. A typical 
example of an isometry without fixed points are the translational 
symmetries on tori. On the other hand, rotational isometries
have fixed points. At the fixed point $k^2=0$. It follows from 
(\ref{dual1}) (using $k^2 = g_{xx}$) that the dual geometry
appears to have a singularity at the fixed point. 

Taking the curvature of the spacetime to be small in string 
units (which is required for consistency for strings propagating
in a background that only solves the lowest order beta functions)
we see that we may approximate the vicinity of the fixed point 
by flat space. In adapted coordinates, which are just polar 
coordinates, the isometry direction being the angular coordinate,
we have
\be \label{flat}
ds^2=dr^2+r^2 d\theta^2.
\ee
Dualizing along $\theta$ we obtain
\be \label{flatd}
ds^2=dr^2+{1 \over r^{2}} d\theta^2, \qquad \phi=- \half \ln r^2.
\ee
So indeed the fixed point of the isometry, i.e. $r=0$, becomes a singular
point after the duality transformation.  Since the curvature now 
diverges at $r=0$ we cannot trust the (first order in $\a'$)
sigma model analysis. A more careful conformal field theory analysis\cite{AAB}
shows that the duality yields an exact equivalence but the operator mapping 
includes all orders in $\a'$. We can read this results as follows:
All order $\a'$ corrections resolve the singularity present 
in the spacetime described by (\ref{flatd}) yielding an exact non-singular 
conformal field theory. 

Studies of T-duality along a rotational 
isometry can be found in \cite{Bak1,BaSf,GHM}. 

\newpage
\section{Brane solutions} \label{low}

String theory has a mass gap of order $1/l_s$. At low enough energies
only the massless fields are relevant. We can decouple
the massive modes by sending $\a' \to 0$ (so the mass of the massive
modes goes to infinity). The interactions of the massless fields
are described by an effective action. For IIA and IIB superstring theories
the low energy theory is IIA and IIB supergravity, respectively. 
We have seen that in type II string theories there exist dynamical 
objects other than strings, namely D-branes, and solitonic 
branes. For each of these objects there is a corresponding 
solution of the low energy supergravity. The purpose of this 
section is to describe these solutions. For reviews 
see \cite{DKL,stelle,Youm}.

The relevant part of the supergravity action, in the string frame, is
\footnote{
There are several other bosonic terms in the action. These terms 
are not relevant for the solutions (\ref{pbrane}) 
since in these solutions there is only a single antisymmetric tensor 
turned on. We have also omitted all fermionic terms.}
\be\label{actionten}
S = {1\over 128 \p^7 g_s^2 \a'{}^4} 
\int d^{10}x\, \sqrt{-g} \big[ e^{-2\phi}
\big( R + 4(\partial \phi)^2 - {1 \over 12} |H_3|^2 \big) - 
 {1\over 2 (p+2)!}\, |F_{p+2}|^2 \big]
\ee
We use the convention to keep the asymptotic value of 
$\phi$ in Newton's constant ($G_N^{(10)}=8 \p^6 g_s^2 \a'^4$), 
so the asymptotic value of $e^\f$ below 
is equal to 1.\footnote{The field equations are invariant under 
$e^\f \to c e^\f, C^{(p+1)} \to c^{-1} C^{(p+1)}$, where $c$ is 
a constant, so one can change conventions by an appropriate choice of $c$.}

The equations of motion of the above action have solutions that have 
the interpretation of describing the long range field
of fundamental strings (F1), D$p$-branes and solitonic fivebranes
(NS5). These solutions are given by\cite{HorStrom}
\bea\label{pbrane}
&&ds^2_{st} = H_i^{\a} [H_i^{-1} ds^2(\bE^{(p,1)}) + ds^2(\bE^{(9-p)})]\nonu
&&e^\phi = H_i^{\b} \nonu
&&A^{(p+1)}_{01 \cdots p}=H_i^{-1}-1, \ \mbox{``electric''}, \ \
\mbox{or}\  \ 
F_{8-p} = \star dH_i, \ \mbox{``magnetic''}
\eea
where $A^{(p+1)}$ is either the RR potential 
$C^{(p+1)}$, or the NSNS 2-form $B$, depending on the solution.
$\star$ is the Hodge dual of $\bE^{(9-p)}$. 
The subscript $i=\{p, F1, NS5\}$ denotes which solution
(D$p$-brane, fundamental string or solitonic fivebrane, respectively)
we are describing.
In order (\ref{pbrane}) to be a solution $H_i$ must be a harmonic function on
$\bE^{(9-p)}$,
\be
\nabla^2 H_i = 0
\ee 
Let $r$ be the distance from the origin of $\bE^{(9-p)}$.
The choice
\be
H_i = 1 + {Q_i \over r^{(7-p)}}, \qquad p<7
\ee
yields the long-range fields of $N$ infinite parallel planar $p$-branes
near the origin. The constant part was chosen equal to one in order the 
solution to be asymptotically flat. The values of the parameters $\a$ and $\b$ 
for each solution are given in Table 2. In the same table we also 
give the values of the charges $Q_i$. The constant $d_p$ is equal 
to  $d_p=(2 \sqrt{\p})^{5-p} \G({7-p \over 2})$. 

\begin{table}[h]
\begin{center}
\vspace{.2cm}
\begin{tabular}{|c|c|c|c|}
\hline
D$p$-branes & $\a=1/2$ & $\b=(3-p)/4$ & $Q_p=d_p N g_s l_s^{7-p}$ \\ \hline
F1          & $\a=0$   & $\b=-1/2    $ & $Q_{F1}=d_1 N g_s^2 l_s^{6}$ \\ \hline
NS5         & $\a=1$ & $\b=1/2$       & $Q_{NS5}=N l_s^{2}$ \\ \hline
\end{tabular}
\caption{{\it $p$-brane solutions of Type II theories.}}
\end{center}
\end{table}

Apart from these solutions, there are also purely gravitational ones.
There is a solution describing 
the long range field produced by momentum modes 
carried by a string. This is the gravitational wave  solution,
\be \label{wave}
ds^2 = -K^{-1} dt^2 + K(dx_1 - (K^{-1}-1)dt)^2 + dx_2^2 + \cdots + dx_9^2
\ee
where $K=1+Q_K/r^6$ is again a harmonic function and 
$Q_K=d_1 g_s^2 N \a'/R^2$. $R$ is the radius of $x_1$.

Finally, there is a  solution describing a Kaluza-Klein (KK) monopole
(the name originates from the fact that upon dimensional reduction 
over $\psi$ the KK gauge field that one gets is the monopole connection):
\bea \label{KK}
&&ds^2=ds^2(\bE^{(6,1)}) + ds^2_{TN} \nonu
&&ds_{TN}^2=H^{-1}(d\psi + Q_M \cos \theta d \varphi)^2 + H dx^i dx^i, 
\qquad i=1,2,3 \nonu
&&H=1+{Q_M \over r}, \qquad r^2=x_1^2 +x_2^2 + x_3^2
\eea
where $TN$ stands for Taub-NUT, $\theta$ and $\psi$ are the angular 
coordinates of $x_1, x_2, x_3$, 
$Q_M=N R/2$, $N$ is the number of coincident monopoles 
and $R$ is the radius of $\psi$.

S-duality leaves invariant the action in the Einstein frame.
To reach the Einstein frame we need to do the Weyl rescaling 
$g_E = e^{-\f/2} g_{st}$. Using the fact that under S-duality 
$\phi \to -\phi$ (and $g_s \to 1/g_s$) 
we get $g_{\m \n} \to e^{-\f} g_{\m \n}$. 
The compactification radii are measured using 
the string metric. So, they change under S-duality. 
One can take care of this by changing the string scale,
$\a' \to \a' g_s$. 
We therefore get the following S-duality transformation rules
\bea
&&\phi \to -\phi \ \ (g_s \to 1/g_s), \qquad \a' \to \a' g_s \nonu
&& g_{\m \n} \to e^{-\f} g_{\m \n}, \qquad 
B_{\m \n} \leftrightarrow C^{(2)}_{\m \n}  
\eea
With these conventions Newton's constant, $G_N^{(10)}=8 \pi^6 g_s^2 \a'^4$,
is invariant under S-duality.

T-duality acts as in (\ref{dual1}) in the NSNS sector. In particular, 
dualization along a coordinate of radius $R$ yields 
\be
R \to {\a' \over R}, \qquad g_s \to g_s {l_s \over R}
\ee
For the RR fields we get\cite{BHO} 
\bea
&&C_{\m_1 \cdots \m_{p+1}} \to C_{\m_1 \cdots \m_{p+1}x}, \qquad
x \not\in \{x_{\m_1}, \cdots, x_{\m_{p+1}}\} \nonu
&&C_{x \m_1\cdots \m_{p+1}} \to C_{\m_1 \cdots \m_{p+1}}
\eea
depending on whether we dualize along a coordinate transverse or parallel 
to the brane.

It is easy to see that the values of the charges $Q_i$ 
are consistent with dualities.
For instance, under S-duality: $Q_{NS5}=N \a' \leftrightarrow N g_s \a' =Q_5$.
Actually, dualities determine 
both the value of Newton's constant and the charges (including
the numerical coefficients) \cite{malda1}:
The mass $M$ of an object can be 
calculated from the deviation of the Einstein
metric from the flat metric at infinity. In particular\cite{coef},
\be 
g_{E,00}={16 \p G_N^{(d)} M \over (d-2) \o_{d-2}} {1 \over r^{d-3}}
\ee
where $\o_{d}=2 {\p^{(d+1)/2}\over \G({d+1 \over 2})}$ 
is the volume of the unit 
sphere $S^d$. Completely wrapping a given brane on torus and dimensionally
reducing we get a spacetime metric in $d=10-p$ dimensions,
\be 
ds^2_{E,d}=-H^{-{d-3\over d-2}} dt^2 + H^{{1 \over d-2}}ds^2(\bE^{(d-1)})
\ee 
This result is obtained by using the dimensional reduction 
rules\cite{dimred}
\be \label{dred}
ds_{E,d}^2 = e^{-{4 \over d-2}\f_d} ds_{st}^2, \qquad
e^{-2 \f_d} = e^{-2 \f} \sqrt{det g_{int}}
\ee
where $g_{int}$ is the component of the metric in the directions we 
dimensionally reduce. If $H=1 + c^{(d)}/r^{d-3}$ then,
\be \label{cd}
c^{(d)}={16 \pi G_N^{(d)} M \over (d-3) \o_{d-2}}.
\ee 
The mass $M$ appearing in this formula is the same as the mass measured
in the string frame since we used the convention to leave 
a factor of $g_s^2$ in Newton's constant. These masses can be easily 
obtained by U-duality.
Knowledge of one of the coefficients in (\ref{cd}) is sufficient to determine 
$G_N$ and therefore all other coefficients as well. In \cite{malda1}
the value of $c_{NS5}$ was determined from the Dirac quantization condition.
Perhaps the simplest way to proceed is to observe that the coefficient in 
the harmonic function of the KK monopole is fixed by requiring that 
the solution is non-singular.

All these solutions are BPS solution and preserve half of maximal
supersymmetry. 
This implies that certain quantities do not renormalize.
Let us sketch the argument. The supersymmetry algebra has the form
\be \label{susy}
\{ Q_\a, Q_\b \} \sim (C \G^\m)_{\a \b} P_\m 
+ (C\G^{\m_1 \cdots \m_p})_{\a \b} Z^{(p)}_{\m_1 \cdots \m_p}
\ee
where $C$ the charge conjugation matrix, $Q_\a$ are the supercharges, 
$P_\m$ is the momentum generator, and 
$Z^{(p)}$ are central charges. 
These are the charges carried by $p$ branes. 

Taking the expectation value of (\ref{susy}) between a physical state
$|A\>$ and going to the rest frame we get
\be \label{BPS1}
\<A|\{ Q_\a, Q_\b \}|A\> = (M^A - c |Z|)_{\a \b} \geq 0
\ee
where $M^A_{\a \b}$ is the mass matrix, $c$ is a constant, and we used the 
fact that $\{ Q_\a, Q_\b \}$ is a positive definite matrix.

If the matrix in the right hand side 
has no zero eigenvalues, then one can take suitable linear combinations
of the supercharges so that the superalgebra takes the form
of fermionic oscillator algebra. Then half of oscillators
can be regarded as creation and half as annihilation operators.
This means that a supermultiplet contains $2^{16}$ states.

If the matrix in the right hand side of (\ref{BPS1}) has a zero 
eigenvalue (so the mass is proportional to the charge, $M=c |Z|$,
i.e. we have a BPS state)
then some of the generators annihilate the state.
The remaining supercharges can again be divided into half creation and half 
annihilation operators.  Thus, the BPS supermultiplet is a short multiplet.
For 1/2 supersymmetric states, such as the branes we have been discussing,
this means that we have $2^8$ states instead of $2^{16}$.

If we vary adiabatically the parameters of theory (i.e. no
phase transition) the number of states cannot change abruptly,
so the number of BPS states remains invariant and
the mass/charge relation does not renormalize\cite{witOl}.
(Here we also assume that we do not cross curves of marginal stability).

\subsection{M-branes}

We briefly describe the connection of the brane solutions described 
in the previous section to M-theory. M-theory at low-energies
is described by eleven dimensional supergravity. The bosonic
field content of eleven dimensional supergravity
consists of a metric, $G_{MN}$, and a three-form
antisymmetric tensor, $A_{MNP}$. We therefore expect that this
theory has solutions describing extended objects coupled 
``electrically'' and ``magnetically'' to $A_{MNP}$. Indeed,
one finds a 2-brane solution, M2, and a fivebrane solution,
M5\cite{DuSt,guven}. The explicit form of the solution is as in (\ref{pbrane}),
with $\a_{M2}=1/3$ for the M2 and $\a_{M5}=2/3$ for the M5
(there is no dilaton field in 11$d$ supergravity, so $\b=0$). 
In addition, we have the purely gravitational solutions
describing traveling waves and KK monopoles.

From the solutions of eleven dimensional supergravity
one can obtain the solution of IIA 
supergravity upon dimensional reduction. The Kaluza-Klein ansatz for the 
bosonic fields leading to the string frame 10$d$ metric is 
\bea
&&ds_{11}^2=e^{-{2 \over 3} \f(x)} g_{\m \n} dx^{\m} dx^{\n} +  
e^{{4 \over 3} \f(x)} (dx_{11} + C^{(1)}_\m dx^\m)^2 \nonu
&&A=C^{(3)} + B \wedge dx_{11}
\eea
where $B$ is the NSNS antisymmetric tensor and $C^{(1)}$ and 
$C^{(3)}$ are the RR antisymmetric tensors of IIA theory.
Dimensionally reducing the M-branes along a worldvolume or 
a transverse direction one obtains all solution of IIA as follows:

\begin{tabular}{ccccccccccccc}
$11d$ SUGRA &\ &\ & W &\ &\ & M2&\ &\ & M5&\ &\ & KK \\
\ &\ &\qquad\swa&\dar&\ &\qquad\swa&\dar&\ &\qquad\swa&\dar&\ 
&\qquad\swa&\dar  \\
IIA SUGRA   &\ & D0 & W &\ & F1 & D2 & \ & D4 & NS5 &\ & D6 & KK
\end{tabular}  

\subsection{Intersection rules}

In the previous section we described brane solutions of supergravity theories. 
These solutions can be used as 
building blocks in order to construct new solutions
\cite{PT,Tse1,GKT,ts1,EREJS,Arefeva,AEH,ohta1,Ber,ohta2,ETT} (for a review
see \cite{Gaunt}). The new solutions
can be interpreted as intersecting (or in some cases overlapping) 
branes. In order to obtain a supersymmetric solution only certain 
intersections are allowed. 

The intersection rules are as follows: \newline
One superimposes the single brane solutions using the 
rule that all pairwise intersections should belong to a set 
of allowed intersections. If all harmonic functions are taken 
to depend on the overall transverse directions (i.e. the directions
transverse to all branes) we are dealing with a ``standard'' intersection. 
Otherwise the intersection will be called ``non-standard''.  
In $D=11$ there are three standard intersections,
$(0|\,2\perp 2)$\footnote{
The notation $(q|\,p_1 \perp p_2)$ denotes a $p_1$-brane intersecting
with a $p_2$-brane over a $q$-brane.},
$(1|\,2\perp 5)$ and $(3|\,5\perp 5)$ \cite{Tse1,GKT,AEH},
and one non-standard $(1|5\perp5)$\cite{khuri,GKT}.
In the latter intersection the harmonic functions depend on the 
relative transverse directions (i.e. the directions which are 
worldvolume coordinates of the one but transverse coordinates of the 
other fivebrane). 
In addition, one can add a wave solution along a common string. 
The intersection rules in ten dimensions can be derived from these
by dimensional reduction plus T and S-duality. We collect 
the standard and non-standard intersection rules in the table below.
(For intersections rules involving KK monopoles see \cite{Ber}).
When both standard and non-standard 
intersection rules are used (as for instance in the solutions of \cite{BPS3}), 
one has to specify which coordinates each harmonic function depends on.
This is usually clear by inspection of the intersection, but it can also be
further verified by looking at the field equation(s)
for the antisymmetric tensor field(s).

\begin{table}[h]
\begin{center}
\vspace{.2cm}
\begin{tabular}{|c|c|c|}
\hline
          & standard            & non-standard       \\ \hline\hline
$D=11$    & $(0|M2\perp M2)$    &                    \\ \hline
          & $(1|M2\perp M5)$    &                    \\ \hline
          & $(3|M5\perp M5)$    & $(1|M5\perp M5)$   \\ \hline\hline
$D=10$    & $(\half(p+q-4)|Dp\perp Dq)$ & $(\half(p+q-8)|Dp\perp Dq)$\\ \hline
          & $(1|F1\perp NS5)$   &                    \\ \hline
          & $(3|NS5\perp NS5)$  & $(1|NS5\perp NS5)$ \\ \hline
          & $(0|F1\perp Dp)$    &                    \\ \hline
          & $(p-1|NS5\perp Dp)$ & $(p-3|NS5\perp Dp)$\\ \hline
\end{tabular}
\caption{{\it Standard and non-standard intersections in ten and eleven
 dimensions.}}
\end{center}
\end{table}

There is a simple algorithm  which leads to 
non-extreme version of a given supersymmetric solution
(constructed according to standard intersection rules)\cite{ts1}.
We will give these rules for M-brane intersections. This is sufficient
as dimensional reduction and duality produce all standard intersections
of type II branes. It consists of the following steps:

(1) Make the following replacements in the $d$-dimensional transverse 
spacetime part of the metric:
\begin{equation}
dt^2  \to   f(r) dt^2 \ , \qquad 
dx_{1}^2 + \cdots  +dx_{d-1}^2 
\to f^{-1} (r) dr^2 + r^2 d \Omega^2_{d-2}\ , 
\qquad f(r) = 1 - {\mu^{d-3} \over  r^{d-3}} \ , 
\label{ii}
\end{equation}
and use the following harmonic functions,
\bea
&&H_T = 1 + {\cq_T\over r^{d-3}} \ , \qquad
\cq_T= \mu^{d-3} \sinh^2 \a_T \ , \nonu
&&H_F = 1 + {\cq_F\over r^{d-3}} \ , \qquad
\cq_F= \mu^{d-3} \sinh^2 \a_F \ , 
\eea
for the constituent two-branes and five-branes, respectively.

(2) In the expression for the field strength  ${\cal F}_4$ of the
three-form field make the following replacements: 
\bea
H^\prime_T{}^{-1} \to 
H^\prime_T{}^{-1} = 1- {Q_T \over r^{d-3}} H_T^{-1}, \qquad
&&Q_T =\mu^{d-3} \sinh \a_T \cosh \a_T\ , \nonu
H_F \to H'_F = 1 + {Q_F \over r^{d-3}}  \qquad
&&Q_F =\mu^{d-3} \sinh \a_F \cosh \a_F\ , 
\eea
in the ``electric''  (two-brane) part, and in the ``magnetic''  
(five-brane) part, respectively. In the extreme limit 
$\mu \to 0 ,\  \a_F \to \infty$, and $\a_T \to \infty$,
while the charges $Q_F$ and $Q_T$ are  kept fixed. In
this case $\cq_F=Q_F$ and  $\cq_T =Q_T$, so that  $H^\prime_T=H_T$.
The form of  ${\cal F}_4$ and the actual value of its ``magnetic''
part does not  change compared to the extreme limit.

(3) In the case there is a common string along some  direction $x$, 
one can add momentum along $x$. Then 
\be \label{nwave}
- f(r) dt^2 + dx^2 \to  
- K^{-1} (r) f(r) dt^2 + K(r) \left(dx  - [K^\prime{}^{-1} (r) -1] dt \right)^2
\end{equation}
where
\bea
K= 1 + {\cq_K \over r^{d-3}} \ , \qquad 
&&\cq_K = \mu^{d-3} \sinh^2 \a_K \ , \nonu
K^\prime{}^{-1}  = 1- {Q_K \over r^{d-3}}  K^{-1} \ , \qquad
&&Q_K = \mu^{d-3}  \sinh \a_K \cosh \a_K \ . 
\eea
In the extreme limit 
$\mu\to 0, \ \a_K\to \infty$, the charge $Q_K$ is held fixed, $K={K^\prime}$ 
and  thus the metric (\ref{nwave}) becomes
$ dudv + (K-1)du^2,$ where   $u,v=x \pm  t$. 

\newpage
\section{Black holes in string theory}

Black holes arise in string theory as solutions of the corresponding 
low-energy supergravity theory. String theory lives in 10 dimensions (or 11 
from the M-theory perspective). Suppose the theory is compactified
on a compact manifold down to $d$ spacetime dimensions. 
Branes wrapped in the compact
dimensions will look like pointlike objects in the $d$-dimensional 
spacetime. So, the idea is to construct a configuration
of intersecting wrapped branes which upon dimensional
reduction yields a black hole spacetime.
If the brane intersection is supersymmetric 
then the black hole will be extremal supersymmetric black hole. 
On the other hand, non-extremal intersections yield non-extremal black holes.

In general, the regime of the parameter space in which supergravity 
is valid is different from the regime in which weakly coupled
string theory is valid. Thus, although we know that a given 
brane configuration becomes a black hole when we go from 
weak to strong coupling, it would seem difficult to 
extract information about the black hole from this fact. 

For supersymmetric black holes, however, the BPS property 
of the states allows one to learn certain things about 
black holes from the weakly coupled D-brane system.
For example, one can count the number of states at 
weak coupling and extrapolate the result to the black hole 
phase. In this way, one derives the Bekenstein-Hawking
entropy formula (including the precise numerical coefficient) 
for this class of black holes\cite{vastro,maldaCal}.
We will review this calculation in section \ref{Extr-D}.

In the absence of supersymmetry, we cannot in general
follow the states from weak to strong coupling. 
However, one could still obtain some qualitative understanding
of the black hole entropy. On general grounds, one might expect that
the transition from weakly coupled strings to black holes
happens when the string scale becomes approximately
equal to the Schwarzschild radius (or more generally to the curvature radius
at the horizon). This point is called the 
correspondence point. Demanding that the mass and the 
all other charges of the two different configurations 
match, one obtains that the entropies also match
\cite{hor3}. These considerations correctly provide
the dependence of the entropy on the mass and the 
other charges, but the numerical coefficient in the 
Bekenstein-Hawking entropy formula remains undetermined. 

In \cite{KS} a different approach was initiated. Instead of trying to 
determine the physics of black holes using the fact that at weak 
coupling they become a set of D-branes, the symmetries of M-theory are
used in order to map the black hole configuration to another black hole
configuration. Since the U-duality group involves strong/weak transitions
one does not, in general, have control over the microscopic
states that make up a generic configuration. We will see, however,
that the situation is better when it comes to black holes!
U-duality maps black holes to black holes with the same 
thermodynamic characteristics, i.e. the entropy
and the temperature remain invariant. This implies 
that the number of microstates that make up the black hole
configuration remains the same. Notice that to reach this conclusion
we did not use supersymmetry, but the fact that the area of the horizon
of a black hole (divided by Newton's constant) tell us how many degree
of freedom the black hole contains. We discuss this approach in
section \ref{BTZs}.

The effect of the U-duality transformations described in section \ref{BTZs}
is to remove the constant
part from certain harmonic functions (and also change the values of some 
moduli). One can achieve a similar result 
by taking the low-energy limit $\a' \to 0$ while keeping fixed the 
masses of strings stretched between different D-branes.
Considerations involving this limit lead to the 
adS/CFT correspondence\cite{malda}.
This will be discussed in section \ref{nHor}.

\subsection{Extremal black holes and the D-brane counting} \label{Extr-D}

We will analyze five dimensional black holes. Four dimensional
ones \cite{4dBH} can be analyzed in a completely 
analogous manner\cite{MalStro,JKM}.
Rotating black holes have been discussed in \cite{rot1,rot2,rot3}.

\subsubsection{5$d$ Extremal Black Holes} \label{extr}

To study extremal charged five dimensional black holes we 
build a configuration of intersecting branes using the supersymmetric 
intersection rules. In particular, we
consider the configuration of $N_5$ D5-brane wrapped in $x_1, \ldots, x_5$,
$N_1$ D1-brane wrapped in $x_1$, 
with $N_K$ momentum modes along $x_1$. The coordinates
$x_i, i=1,\ldots, 5$ are taken periodic with periods $R_i$.
Explicitly, the spacetime fields are 
\bea \label{extrBH}
ds^2=H_1^{1/2} H^{1/2}_5 && \left[ H_1^{-1} H_5^{-1} 
\left(-K^{-1}dt^2 + K (dx_1 - (K^{-1}-1)dt)^2\right) \right. \nonu
&&\left.+H_5^{-1}(dx_2^2 + \cdots + dx_5^2)
+ dx_6^2 + \cdots + dx_9^2 \right]
\eea
and
\bea
&&e^{-2 \f} = H_1^{-1} H_5, \qquad 
C^{(2)}_{01}=H_1^{-1}-1 \nonu
&&H_{ijk}=\half \e_{ijkl} \pa_l H_5, \qquad i,j,k,l=6, \ldots,9 \\
&&r^2=x_6^2 + \cdots +x_9^2
\eea
The harmonic functions are equal to 
\bea \label{extrHar}
&&H_1=1+{Q_1 \over r^2}, \qquad Q_1={N_1 g_s \a'{}^3 \over V} 
\nonu
&&H_5=1+{Q_5 \over r^2}, \qquad Q_5=N_5 g_s \a' \nonu
&&K=1+{Q_K \over r^2}, \qquad Q_K={N_K g_s^2 \a'{}^4 \over R_1^2 V} 
\eea
where $V=R_2 R_3 R_4 R_5$ and 
the charges have been calculated using (\ref{cd}).

Upon dimensional reduction over the periodic coordinates 
$x_1, \ldots, x_5$, using (\ref{dred}), we obtain 
\be
ds^2_{E,5} = \l^{-2/3} dt^2 + \l^{1/3}(dr^2 + r^2 \d \O_3^2)
\ee
where 
\be
\l = H_1 H_5 K =(1+{Q_1 \over r^2}) (1+{Q_5 \over r^2}) (1+{Q_K \over r^2})
\ee
This is an extremal charged black hole. The horizon is located at $r=0$.
The area of the horizon and the five dimensional Newton's constant
are equal to 
\bea
&&A_5=(r^2 \l^{1/3})^{3/2} \big|_{r=0} \w_3 = \sqrt{Q_1 Q_2 Q_K} (2 \p^2) \nonu
&&G_N^{(5)}={G^{(10)}_N \over (2 \p)^5 R_1 V}
\eea
Therefore, the entropy is equal to 
\be \label{extS}
S={A_5 \over 4 G_5} = 2 \pi \sqrt{N_1 N_5 N_K}
\ee

For the supergravity to be valid we need to suppress string loops and 
$\a'$ corrections. We suppress string loops by sending $g_s \to 0$,
while keeping the charges $Q_i$ fixed. These charges are
the characteristic scales of the system. In order to suppress 
$\a'$ corrections they should be much larger than the string scale,
\be
Q_1,\ Q_5,\ Q_K \gg \a'
\ee
Taking the compactification radii to be of order $l_s$ we obtain
\be \label{val1}
g_s N_1,\ g_s N_5,\ g_s^2 N_K \gg 1
\ee
This means that $N_K \gg N_1 \sim N_5 \gg 1$.

\subsubsection{D-brane counting} \label{Dbr}

We now turn to the weak-coupling D-brane configuration in order
to compute the D-brane entropy.     
Counting the degeneracy of D-brane states translates into 
the question of counting BPS states in the D-brane
worldvolume theory\cite{witten1,sen,vaf1,vaf2,vaf3}.
For the system we are interested in, and taking the 
torus $T^4$ in the relative transverse directions to be small,
$R_2, R_3, R_4, R_5 \ll R_1$, the relevant worldvolume theory is 
1+1 dimensional. This theory is the infrared
limit of the Higgs branch of the 1+1 gauge theory, and it has been 
argued to be a deformation of the 
supersymmetric $\cn=(4,4)$ sigma model with target space 
$(T^4)^{N_1 N_5}/S^{N_1 N_5}$\cite{vaf1}. Since 
$(T^4)^{N_1 N_5}/S^{N_1 N_5}$ is a hyperkaehler
manifold of dimension $4 N_1 N_5$, the sigma model has central charge 
equal to $6 N_1 N_5$. This is the central charge of  
$4 N_1 N_5$ bosonic and fermionic degrees of freedom (since 
scalars contribute 1 and fermions 1/2 to the central charge).
Roughly, these degrees of freedom are the ones describing the 
motion of the D1 brane inside the D5 brane. For details we refer 
to \cite{malda1}.

In the worldvolume
theory we get that the right movers are in their ground state
and the left movers carry $N_K$ momentum modes.
Thus, the degeneracy 
of the D-brane system is given by the degeneracy of the 
conformal field theory of central charge $c=6 N_1 N_5$ 
at level $N_K$. For a unitary conformal field theory the degeneracy 
is given by Cardy's formula\cite{cardy} 
\be \label{deg}
d(c, N_K) \sim \exp (2 \p \sqrt{{c \over 6} N_K})
\ee
Therefore, the entropy is equal to 
\be
S = \log d(c, N_K) = 2 \p \sqrt{N_1 N_5 N_K}
\ee
This is in exact agreement with (\ref{extS}).

Let us now inspect the regime of validity of the D-brane picture.
Open string diagrams pick up a factor $g_s N_{1,5}$ because the 
open string coupling constant is $g_s$ and there are $N_{1,5}$ branes
where the string can end (or equivalently one should 
sum over the Chan-Paton factors). Processes involving momenta involve
a factor $g_s^2 N_K$\cite{SDP}. Therefore,
conventional D-brane perturbation theory is good when
\begin{equation}\label{val2}
g_s N_1,\ g_s N_5,\ g_s^2 N_K \ll 1 \Rightarrow Q_1,\ Q_5,\ Q_K \ll \a',
\end{equation}
which is precisely the opposite regime to (\ref{val1}) where the
classical supergravity solution is good.  The D-brane/string
perturbation theory and black hole regimes are thus complementary.
This feature is related to open-closed string duality.
Due to supersymmetry, however, one can extrapolate results 
obtained in the D-brane phase to the black hole phase.

\subsection{Non-extremal black holes and the BTZ black hole}\label{BTZs}

In this section we review the approach of \cite{KS}. 
The idea is to use U-dualities in order to connect 
higher dimensional black holes to lower dimensional ones.
Such ideas also appeared in \cite{hyun}. 
The U-duality transformation essentially maps the initial black hole
to its near-horizon region (but Schwarzschild black holes are also 
included as a limiting case). In particular, four and five dimensional
black holes are mapped to the three dimensional BTZ black hole. 
The U-duality group of string (or M) theory on a torus does 
not change the number of non-compact dimensions. 
However, black hole spacetimes always contain an extra timelike
isometry. This extra isometry allows for a duality transformation,
the shift transformation\cite{BPS1}, that yields 
trans-dimensional transformations. A thorough 
discussion (that includes global issues) of the shift transformation
is given in section \ref{shiftse}.

\subsubsection{U-duality and entropy}

Let us discuss whether one can use U-duality in order to 
infer a state counting for a given black hole from the counting of a
U-dual configuration. The U-duality group is conjectured (and widely 
believed) to be an exact symmetry of M-theory.
This symmetry, however, is spontaneously broken by the vacuum.
The vacua of M-theory (compactified on some manifold) are parametrized 
by a set constants. These constants are expectation values 
of scalar fields arising from the compactification. U-duality 
acts on these scalars, so it transforms one vacuum to another.
Therefore, from a state on a given vacuum 
one can deduce by U-duality the existence of another state in a new vacuum. 
Since the U-duality group contains S-duality which is strong/weak
coupling duality, one cannot in general continue the new state back 
to the original vacuum, unless this state is protected from quantum 
corrections. States with this property are BPS states.
Therefore, the spectrum of BPS states is invariant under U-duality 
transformations. This implies in particular that if we want to 
count the number of states that make up an extremal supersymmetric
black hole, we may use any U-duality configuration. Indeed, 
the entropy formula for extremal black holes is U-duality 
invariant\cite{FerKal,HMS,CvHu,FeMa}. 

The question is whether it is justified to use U-duality in more 
general context. A remarkable fact about 
S and T duality transformations is that they leave invariant both the entropy 
and the temperature of black holes connected by S and T transformations. 
For S-duality this follows from the fact that S-duality leaves invariant the 
Einstein metric. For T-duality, this has been shown in \cite{hor1}. 
We review this argument here.

Consider a black hole solution 
with a timelike isometry $\pa/\pa t$, a compact spacelike isometry 
$\pa/\pa x$, and a NSNS 2-form $B$ turned on.
Smoothness near the horizon requires\cite{hor1} that 
the $B_{tx}$ vanishes at the horizon.
In order the T-dual geometry to also be smooth (i.e.
the dual 2-form to vanish at the horizon) we require in addition
that $A_x=0$ at the horizon (see (\ref{KKmetric})-(\ref{dual2})).
(This can always be achieved by a coordinate transformation.)
RR potentials that can be turned into $B_{xt}$ by dualities
are also required to vanish at the horizon.
 
Let us first discuss the entropy. In $d$ dimensions the Einstein metric
is given by (see (\ref{dred})),
\be \label{Einme}
ds_E^2 = e^{-4 \f/(d-2)}[g_{xx}(dx + A_i dx^i)^2 + \bar{g}_{ij} dx^i dx^j]
\ee
The metric induced on the horizon is of the same form but with $i,j$
taking values only over the $d-3$ angular variables.
Therefore, the area is equal to
\be
A_d=\int \sqrt{(e^{-4 \f/(d-2)})^{d-2} g_{xx} \det \bar{g}} = 
\int e^{-2 \f} \sqrt{g_{xx}} \sqrt{\det \bar{g}}
\ee
One may check that $e^{-2 \f} \sqrt{g_{xx}}$ is a T-duality invariant
combination (and $\bar{g}$ was invariant to start with). 
Therefore, the entropy of black holes is T-duality invariant.

Let us also note that the entropy formula is 
invariant under dimensional reduction
\be
S={A_{10} \over 4 G_N^{(10)}} = {A_{d} \over 4 G_N^{(d)}}
\ee
since $A_{d}=A_{10}/V_{10-d}$ and $G_N^{(d)}=G_N^{(10)}/V_{10-d}$,
where $V_{10-d}$ is the volume of the compactification space.

We now turn to the discussion of the behavior of the Hawking 
temperature under duality transformations. Perhaps the simplest
way to compute the Hawking temperature is to analytically continue
to Euclidean space by taking $t \to \tau=-i t$. 
The black hole spacetime becomes then a 
non-singular Riemannian manifold provided that the Euclidean 
time is periodically identified with period the inverse 
Hawking temperature. Suppose that the horizon is at $r=\m$. One can calculate
the temperature to be equal to (we assume that the event horizon
is non-degenerate)
\be \label{THaw}
T_H = {\pa_r g_{\t \t} \over 4 \p \sqrt{g_{\t \t} g_{rr}}} \Bigg|_{r=\m}
\ee
It follows by inspection that the Hawking temperature is invariant 
under non-singular Weyl rescaling. Hence, it does not make any difference
whether we consider the Einstein or the string frame. 
We choose to work with the string frame. From (\ref{KKmetric}) we get
\be
g_{\t \t} = \bar{g}_{\t \t} + g_{xx} A_{\t} A_{\t}, \qquad
g_{rr}=\bar{g}_{rr} + g_{xx} A_r A_r
\ee
Assuming that $A_r$ is finite at the horizon (in all case we will 
consider $A_r=0$), and using $g_{xx}|_{r=\m}=A_{\t}|_{r=\m}=0$ we obtain
\be
T_H = 
{\pa_r \bar{g}_{\t \t} \over 4 \p 
\sqrt{\bar{g}_{\t \t} \bar{g}_{rr}}} \Bigg|_{r=\m}
\ee
which is manifestly T-duality invariant. 

Therefore, an arbitrary combination of S and T transformations 
will lead to a black hole solution with the same entropy
and temperature as the original one. 
This implies that black holes 
connected by U-duality transformations have the same number 
of microstates. This is somewhat surprising since 
for non-supersymmetric black holes we cannot 
follow the states during U-duality transformations.
As we move from one configuration 
to a U-dual one, some states may disappear. However, an equal number 
of states has to appear, since the final configuration has the same 
entropy. We do not have a microscopic derivation
of this fact. We believe that such derivation will be 
an important step towards further understanding of black holes.

A general U-duality transformation may involve strong/weak transitions.
The U-duality transformations, however, that we will use below 
do not involve such strong/weak 
transitions. Actually we shall exclusively be in the black hole 
phase. We will only consider transformations, call them
$U_T$, that are connected to T-dualities by a similarity transformation
\be \label{udual}
U_T = U^{-1} T U
\ee
where $U$ denotes a generic U-duality transformation and $T$ a 
sequence of two T-duality transformations (so $U_T$ acts within 
the same theory).

\subsubsection{The shift transformation} \label{shiftse}

As we have discussed, we construct black holes configurations 
using appropriate non-extremal intersections of 
extremal branes. These configurations are solutions of the 
field equations provided the various harmonic functions $H_i$
appearing in the solution satisfy Laplace's equations,
\be
\nabla^2 H_i =0, 
\ee
where $\nabla$ is the Laplacian in the overall transverse space.
The constant part of the harmonic function is usually 
set to one in order the solution to be asymptotically
flat. Clearly, up to normalization, the only other choice is to 
set this constant to zero. This choice has the dramatic effect of
changing the asymptotics of the solution. We will see, however,
that there is a duality transformation, the shift transformation,
that removes the one from the harmonic function. This duality 
transformation has been appeared in the past in various contexts 
\cite{hor2,AABL,Bak1,hyun,BPS1,KS,berg,CLLPST}.

Consider the solution describing a non-extremal fundamental string
in $d+1$ dimensions
\bea \label{sol1}
ds^2&=& H^{-1}(r) (-f(r) dt^2 + dx_1^2) + f^{-1}(r) dr^2 + 
r^2 d \O_{d-2}^2 \nonu
B_{t x_1} &=& H^\prime{}^{-1} -1 + \tanh \a \nonu
e^{-2 \f}& =& H
\eea
The coordinate $x_1$ is periodic with period $R_1$.
The various harmonic functions are equal to  
\bea
H= 1 +{\m^{d-3} \sinh^2 \a \over r^{d-3}},\qquad
H^\prime{}^{-1}=1 - {\m^{d-3} \sinh \a \cosh \a \over r^{d-3}} H^{-1},\qquad
f = 1 - {\m^{d-3} \over r^{d-3}}
\eea
The constant part of the antisymmetric tensor $B_{t x_1}$ is fixed by 
the requirement that $B_{t x_1}$ vanishes at the horizon.
This is required by regularity\cite{hor1}, as described in the
previous section. The entropy and the temperature
are given by
\be \label{area}
S = {1 \over 4 G_N^{(d+1)}} 2 \p R_1 \cosh \a \m^{d-2} \w_{d-2}, \qquad
T_H={(d-3) \over 4 \p \m \cosh \a}.
\ee
Notice that in order to calculate the area one first has to reach the 
Einstein frame.

We now perform the following sequence of T-dualities that we call 
the shift transformation:
\be \label{shift}
shift=T_{{\pa \over \pa t^\prime}} ({\pa \over \pa x_1^\prime})
\circ T_{{\pa \over \pa t}} ({\pa \over \pa x_1}) 
\ee
where
\bea \label{kil}
&&{\pa \over \pa x_1^\prime} =  
-e^{-\a} {\pa \over \pa t} + {1 \over \cosh \a} {\pa \over \pa x_1} \nonu
&&
{\pa \over \pa t'} = \cosh \a {\pa \over \pa t}
\eea
The notation $T_{k_1}(k_2)$ indicates a T-duality transformation 
along the killing vector $k_2$ keeping $k_1$ fixed. 

Let us give the details. After the first T-duality, 
$T_{\pa/\pa t}(\pa/\pa x_1)$, we get a non-extremal wave solution,
\bea
ds^2 &=& -H^{-1}(r) f(r) dt^2 +  H(r) \left(dx_1 - 
(H'{}^{-1}(r)-1+\tanh \a)dt \right)^2 \nonu
&&+ f^{-1}(r) dr^2 + r^2 d \O_{d-2}^2 ~ , \label{wave1} 
\eea
The radius of $x_1$ is now $\a'/R_1$. In addition, $g_s \to l_s/R_1$,
so $G_N^{(d+1)} \to G_N^{(d+1)} \a'/R_1^2$. 
One can check that this solution has the same entropy 
and temperature as the solution in (\ref{sol1}). 

We would like now to dualize along (\ref{kil}). To do this 
we first reach adapted coordinates 
\be \label{transf}
\left(
\begin{array}{c}
t \\
x_1
\end{array}
\right) = 
\left(
\begin{array}{cc}
\cosh \a & - e^{-\a}  \\
0 & {1 \over \cosh \a}
\end{array}
\right)
\left(
\begin{array}{c}
t^\prime \\
x_1^\prime
\end{array}
\right)~ .
\ee
The metric in the new coordinates takes the form (we have dropped the primes)
\be
ds^2 = -\tilde{H}^{-1}(r) f(r) dt^{2} + \tilde{H}(r) 
\left(dx_1 - (\tilde{H}^{-1}(r)-1)dt \right)^2 
+ f^{-1}(r) dr^2 + r^2 d \O_{d-2}^2 ~ ,
\label{wave2}
\ee
where now
\be \label{newha} 
\tilde{H}(r) = \frac{\m^{d-3}}{r^{d-3}} ~ .
\ee 
The radius of $x_1$ also changes to $\cosh \a/R_1$.

Now, that we have reached adapted coordinates we can use (\ref{dual1})
to obtain,
\bea \label{dualst}
ds^2 &=& \tilde{H}^{-1}(r) (-f(r) dt^2 + dx_1^2) + f^{-1}(r) dr^2 
+ r^2 d\O_{d-2}^2 \nonu
B_{\t x_1}&=&\tilde{H}^{-1}-1\nonu
e^{-2 \f} &=& \tilde{H}
\eea
The radius of $x_1$ is now equal to $R_1/\cosh \a$.
In addition, there is a again a change in Newton's constant.
One can calculate the temperature and entropy of this solution.
The result for the entropy is the same in (\ref{area}). 
The temperature is equal to $T_H=(d-3)/4 \p \m$. This differs by a
factor of $\cosh \a$ from (\ref{area}). This is due to the 
fact that the timelike killing vectors $\pa/\pa t$ and $\pa/\pa t'$
differ by a factor of $\cosh \a$ (see (\ref{kil})). 

To summarize, the effect of the shift transformation (\ref{shift})
is to change the solution by removing the constant part of the 
harmonic functions. All the dependence of the metric and the antisymmetric
tensor on the non-extremality angle $\a$ resides in the radius 
of the compact direction which after the shift transformation 
is equal to $R_1/\cosh \a$. In addition,  
$g_s \to g_s/\cosh \a$, so $G_N^{(d+1)} \to G_N^{(d+1)}/\cosh^2 \a$.

The orbits of the killing vector $\pa/\pa x_1^\prime$ are non-compact
since the time coordinate is non-compact. This means that (\ref{sol1})
and (\ref{dualst}) are not equivalent. To make the duality transformation
a symmetry we need to compactify the orbits of the killing vector
$\pa/\pa x_1^\prime$.\footnote{One way to make the orbits
compact is to compactify time with appropriately chosen radius.
It has been argued in \cite{Gibb2} that a spatially wrapped
brane should also be wrapped in time in order to avoid 
conical singularities at the horizon. The two issues may be related.
The time coordinate is naturally compact in Euclidean black holes,
the radius of the time coordinate being the inverse of the Hawking temperature.
One may try to formulate the analysis in the Euclidean framework.
The problem is then that the coordinate transformation (\ref{transf})
is complex.}  
The fact, however, that the entropy and temperature 
of the one black hole can be deduced from the entropy and temperature 
of the other indicates that the two solutions are in the same universality 
class (in a loose sense). 

The norm of the killing vector (\ref{kil}) is 
\be \label{norm}
|\pa /\pa x_1^\prime|^2 = {\m^{d-3} \over r^{d-3}}
\ee
therefore the isometry is spacelike everywhere but it becomes null
at spatial infinity. Let us examine the $(r, x_1)$ part 
of the metric close to spatial infinity.
From (\ref{wave2}) we get 
\be \label{nearfixed}
ds^2_{(r, x_1)} = dr^2 + {\m^{d-3} \over r^{d-3}} dx_1^2 
\ee
For $d=5$, which will be the case in the next section where we discuss 
five dimensional black holes, this is exactly the same metric as in
(\ref{flatd}). This suggest to consider $r$, $x_1$ as polar coordinates 
and the isometry in $x_1$ as a rotational isometry 
with a fixed point at infinity.

\subsubsection{Connection of $5d$ and $4d$ black holes to the BTZ black hole}

We are now ready to use our results to study non-extremal $5d$ and $4d$ black 
holes. We will explicitly work out the case of $5d$ black holes. 
The analysis of $4d$ black holes is completely analogous \cite{KS}.
Four and five dimensional black holes can also be mapped 
by similar operations \cite{hyun,KS,teo} 
to two dimensional black holes\cite{2dBH}.
Let us also note that the manipulations we describe here
cannot connect the BTZ black hole 
to higher than five dimensional black holes \cite{KS}. 
The relation between the near-horizon limit of higher-dimensional
black holes and the BTZ black hole has also been investigated in \cite{satoh}.

The solution we will study is the non-extremal version of 
(\ref{extrBH}). Explicitly, the metric, the dilaton 
and the antisymmetric tensor are given by
\bea \label{10d}
ds_{10}^2 =  
H_{1}^{1/2} H_5^{1/2}&&\left[H_{1}^{-1} H_5^{-1} 
\left(-K^{-1} f dt^2 
+ K \left(d x_1 - (K'{}^{-1} -1)dt\right)^2\right)\right. \nonu
&&\left.+H_5^{-1}(dx_2^2 + \cdots + dx_5^2) + (f^{-1} dr^2 + r^2 d\O_3^2)
\right] ~ ,
\eea 
and
\bea
&&e^{-2 \f} = H_{1}^{-1} H_{5}~ ,~~~~ 
C^{(2)}_{01} = H_{1}'{}^{-1}-1 + \tanh \a_1~ , 
\nonu
&& H_{ijk} = {1 \over 2} \e_{ijkl} \pa_l H_{5}'~ , ~~~~  
i,j,k,l=6, \ldots, 9 ~ ,
\label{100d}\\
&&  f=1 -{\m^2 \over r^2}~ , ~~~~  r^2 = x_6^2 + \cdots + x_9^2 ~ ,\nonumber
\eea
The coordinates $x_i$, $i=1, \ldots, 5$, are assumed to be periodic, 
each with radius $R_i$. 

The various harmonic function are given by 
\bea \label{harm1}
&&K=1 + \frac{\cq_K}{r^2}, \  
K'{}^{-1} = 1 - \frac{Q_K}{r^2} K^{-1}, \ 
\cq_K=\m^2 \sinh^2 \a_K, \ 
Q_K= \m^2 \sinh \a_K \cosh \a_K \nonu
&&H_1=1 + \frac{\cq_1}{r^2}, \  
H'_1{}^{-1}= 1 - \frac{Q_1}{r^2} H_1^{-1}, \ 
\cq_1=\m^2 \sinh^2 \a_1, \ Q_1= \m^2 \sinh \a_1 \cosh \a_1 \nonu
&&H_5=1 + \frac{\cq_5}{r^2}, \  
H'_5= 1 + \frac{Q_5}{r^2}, \ 
\cq_5=\m^2 \sinh^2 \a_5, \ Q_5= \m^2 \sinh \a_5 \cosh \a_5 ~ ,
\eea

Dimensionally reducing in $x_1, x_2, x_3, x_4, x_5$, one gets  
a $5d$ non-extremal black hole, whose metric in the Einstein frame is given by
\be
ds_{E, 5}^2 = - \l^{-2/3} f dt^2 + \l^{1/3} (f^{-1} dr^2 + r^2 d \O_3^2)~ ,
\label{BH5d}
\ee
where 
\be
\l = H_{5} H_1 K = \left(1 + \frac{\cq_{5}}{r^2}\right)\left(1 
+ \frac{\cq_1}{r^2}\right)\left(1 + \frac{\cq_K}{r^2}\right)~ .
\ee
This black hole is charged with respect to the Kaluza-Klein gauge fields 
originating from the antisymmetric tensor fields and the metric. When all 
charges are set equal to zero one obtains the $5d$ Schwarzschild black hole.
The metric (\ref{BH5d}) 
has an outer horizon at $r=\m$ and an inner horizon at $r=0$. 

The Bekenstein--Hawking entropy may easily be calculated to be
\be \label{ent}
S={A_5 \over 4 G_N^{(5)}} = 
{1 \over 4} \frac{(2 \p)^5 R_1 V}{G_N^{(10)}} 
 \m^3 \o_3 \cosh \a_{5} \cosh \a_1 \cosh \a_K~ ,
\ee
where $V=R_2 R_3 R_4 R_5$ is the compactification volume 
in the relative transverse directions,
$\o_3$ is the volume of the unit 3-sphere and 
$G_N^{(5)}$ and $G_N^{(10)}$ are Newton's constant in five and ten dimensions, 
respectively. The temperature is given by
\be
T_H = {1 \over 2 \p \m \cosh \a_1 \cosh \a_5 \cosh \a_K}
\ee

We will now show that one can connect this black hole to the 
BTZ black hole times a 3-sphere 
using transformations of the form (\ref{udual}).
A U-transformation is used to map a given brane to a fundamental string.
The T transformation is the shift transformation (\ref{shift}).

For the case at hand we need to perform the shift transformation 
to the D1 and the D5 brane. The final result is given
by the metric in (\ref{10d}), but with
\be
H_{1} = \frac{\m^2}{r^2}~ ,~~~~~~~~ H_{5} = \frac{\m^2}{r^2}~ ,
\label{hfhs5}
\ee    
and, in addition,
\bea
&& e^{-2 \f}=1~ ,~~~~~~ C^{(2)}_{01}=H_{1}^{-1}-1~ ,
\nonumber \\
&& H_{ijk}={1 \over 2}\e_{ijkl}\pa_l (H_{5}{-}1)~ ~ ,~~~~~
i,j,k,l=6, \ldots, 9
\label{p01h} ~ .
\eea
In addition the compactification volume becomes 
$V \to V/(\cosh \a_1 \cosh \a_5)$ (here, for convenience
in the presentation, we assume that the U-duality 
transformation mapped the D1 and D5 into a fundamental string  
wrapped in one of the relative transverse directions). 
Furthermore,
$G_N^{(10)} \to G_N^{(10)}/(\cosh^2 \a_1 \cosh^2 \a_5)$. 
Notice that the parameters $\a_1$ and $\a_{5}$ associated to the charges
of the original D1 and D5 brane do not appear in the background 
fields anymore.

Dimensionally reducing along $x_2, x_3, x_4, x_5$ we find
\be
ds^2_{E,6}=ds_{BTZ}^2 + l^2 d \O_3^2 ~ ,
\label{6sol}
\ee
where 
\be
ds_{BTZ}^2= -\frac{(\r^2 - \r_+^2)(\r^2 - \r_-^2)}{l^2 \r^2} dt^2
+ \r^2 (d \varphi + \frac{\r_+ \r_-}{l \r^2} dt)^2 + 
\frac{l^2 \r^2}{(\r^2 - \r_+^2)(\r^2 - \r_-^2)} d \r^2 ~ 
\label{dsbtz}
\ee
is the metric of the non-extremal BTZ black hole 
in a space with cosmological constant 
$\L=-1/l^2$, with inner horizon at $\r=\r_-$ and  outer horizon at $\r=\r_+$.  
The mass and the angular momentum of the BTZ black hole are equal to 
\be
M={\r_+^2 + \r_-^2 \over 8 G_N^{(3)} l^2}, \qquad
J={\r_+ \r_- \over 4 G_N^{(3)} l}. 
\ee
In terms of the original variables:
\bea
&& 
l =\m~ ,~~~~~ \varphi={x_1 \over l}~ , ~~~~~ \r^2 = r^2 + l^2 \sinh^2 \a_K~ ,
\nonumber\\
&& \r_+^2 = l^2 \cosh^2 \a_K~ ,~~~~~~ \r_-^2 = l^2 \sinh^2 \a_K ~ .
\label{param}
\eea  
In addition,
\be
\f =0~ ,~~~~~  C^{(0)}_{t \varphi}=(\r^2 - \r_+^2)/l~ ,~~~~~ H =l^2 \e_3~ ,
\label{pbtp}
\ee
where $\e_3$ is the volume form element of the unit 3-sphere.
Therefore, the metric (\ref{6sol}) 
describes a space that is a product of a 3-sphere of radius $l$ and of 
a non-extremal BTZ black hole. 
Notice that the BTZ and the sphere part are completely decoupled. 

We can now calculate the entropy of the resulting black hole.
The area of the horizon is equal to 
\be
A_3= 2 \pi {R_1 \over \m} \m \cosh \a_K~ ,
\ee
whereas Newton's constant is given by 
\be
G_N^{(3)} ={ G^{(10)}_N \over (2 \p)^4 V (\cosh \a_1 \cosh \a_5)
(\m^3 \o_3)}  ~ .
\label{new3}
\ee
It follows that $S=A_3/(4 G_N^{(3)})$ equals (\ref{ent}), i.e. 
the Bekenstein--Hawking entropy of the final configuration is equal to the 
one of the original $5d$ black hole. Notice that the Newton constant in 
(\ref{new3}) contains the parameter $\a_1, \a_{5}$, i.e. carries information on
the charge of the original D1 and D5 brane. The temperature of the BTZ
black hole is equal to 
\be
T_{BTZ} = {\r_+^2 - \r_-^2 \over 2 \p \r_+ l^2}
\ee
Transforming to the original variables we get 
\be
T_{BTZ}={1 \over 2 \p \m \cosh \a_K}=\cosh \a_1 \cosh \a_5 T_H
\ee
precisely as predicted by the duality transformations.

We finish this section by pointing out a remarkable fact:
We have started with the solution (\ref{10d}) of the low-energy supergravity.
This solution is expected to get $\a'$ corrections. Then we used
the T-duality rules (\ref{dual1}) which are also valid  
only to first order in $\a'$.
The final result, however, is valid to all orders in $\a'$!

The fields in (\ref{dsbtz}), (\ref{pbtp}) have their canonical value, 
so that both the BTZ and the sphere part  
are separately exact classical solutions of string theory,\footnote{
For the D1-D5 system that we discuss we obtain a CFT 
describing a D-string.
One gets a fundamental string from the S-dual system of F1-NS5.} i.e. 
there is an exact CFT associated to each of them.
For the BTZ black hole the CFT corresponds to an orbifold of the WZW model
based on $SL(2,\IR)$ \cite{hor2,kaloper,kumar}, 
whereas for $S^3$ and the associated
antisymmetric tensor with field strength $H$, given in (\ref{pbtp}), the 
appropriate CFT description is in terms of the $SO(3)$ WZW model.
The same result also holds in the case of $4d$ black holes\cite{KS}.
This time the black hole is mapped to $BTZ \times S^2$. 
Again all fields are such that there is an exact CFT associated 
to each factor. The one associated with $S^2$ is the monopole 
CFT of \cite{GPSmonopole}.

The situation
seems quite similar to the case described at the end of section \ref{bsec}:
There we had the singular solution (\ref{flatd}) of the lowest order 
in $\a'$ beta function equation which becomes an exact CFT after dualization
with respect to a killing vector whose norm vanishes at spatial infinity.
However, to establish equivalence one needs all order in $\a'$. 

In the case of black holes we have: \newline
The singular black hole spacetime (\ref{10d}) 
that solves the lowest order in $\a'$ beta functions becomes,
after dualization with respect to a killing vector whose
norm vanishes at spatial infinity (plus other dualities),
the BTZ black hole which contains no curvature singularity 
and is an exact CFT. (So, one could argue that 
the original singularity is resolved by $\a'$ corrections).

We find these similarities quite suggestive. However, it is difficult 
to see how one could overcome the problem of the non-compactness of the
orbits of the killing vector in (\ref{kil}). 

\subsection{Low-energy limit and the near-horizon geometry} \label{nHor}

\subsubsection{Near-horizon limit of branes}

We have argued that the physical system describing a
black hole in strong coupling becomes a set of intersecting 
branes in weak coupling. We emphasize that there is only one 
physical system. Its description,
however, in terms of some weakly coupled theory changes as we change 
the parameters of the theory, and furthermore, 
at any given regime of the parameter 
space, there is only one weakly coupled description.

One may view the different descriptions as effective theories
that are adequate to describe the system at specific range 
of the parameter space. As we go  outside this range
new degrees of freedom become important and a new description
takes over. In some cases, however, a given theory may 
still be well-defined for any value of the coupling constant. 
In this case we get a dual description of this theory.

Let us consider $N$ coincident D$p$-branes. At weak coupling they 
have a description as hypersurfaces where string can end.
There is worldvolume theory describing the collective 
coordinates of the brane. The worldvolume 
fields interact among themselves and with the bulk fields.
We would like to consider a limit which decouples the bulk 
gravity but still leaves non-trivial dynamics on the worldvolume.
In low energies gravity decouples. So, we consider the 
limit $\a' \to 0$ , which implies that the gravitation coupling constant,
i.e. Newton's constant, $G_N \sim \a'^4$, also goes to zero.
We want to keep the worldvolume degrees of freedom 
and their interactions. Since the worldvolume dynamics
are governed by open string ending on the D-branes, we keep 
fixed the masses of strings stretched between D-branes as 
we take the limit $\a' \to 0$. In addition, we keep fixed the 
coupling constant of the worldvolume theory, so 
all the worldvolume interactions remain present. 
For $N$ coincident D-branes, the worldvolume theory is 
an $SU(N)$ super Yang-Mills theory (we ignore the center of mass
part). The YM coupling constant
is equal (up to numerical constants) to $g_{YM}^2 \sim g_s (\a')^{(p-3)/2}$. 
Thus we get that the following limit,
\be \label{llimit}
\a' \to 0, \qquad U={r \over \a'}={\rm fixed}, \qquad g_{YM}^2={\rm fixed}
\ee
yields a decoupled theory on the worldvolume.

At strong coupling the D$p$ branes are described by the black 
$p$-brane spacetimes (\ref{pbrane}). Let us consider the 
limit (\ref{llimit}) for this spacetime. One gets that the 
harmonic function becomes,
\be \label{harsh}
H_i \to g_{YM}^2 N (\a')^{-2} U^{p-7}
\ee
The limit (\ref{llimit}) is a near-horizon limit since 
$r=U\a' \to 0$ and there is a horizon at $r=0$.
We see that the effect of the limit (\ref{llimit})
is similar to the effect of the shift 
transformation, namely the one is removed from the harmonic function.
Inserting (\ref{harsh}) back in the metric one 
finds that the spacetime becomes conformal to 
$adS_{p+2} \times S^{8-p}$ \cite{BPS3,BST}
(for M-branes, one gets $adS_4 \times S^7$ for the M2 brane and 
$adS_7 \times S^4$ for the M5 brane \cite{m5}).

Let us now consider the particular case of $N$ coincident D3-branes.
The worldvolume theory 
is $d=4$, $\cn=4$ $SU(N)$ SYM theory. This is a finite unitary theory 
for any value of the its coupling constant. On the other 
hand,  this system has a description as
a black 3-brane at strong coupling. In the limit (\ref{llimit}) we get that 
the spacetime becomes $adS_5 \times S^5$. In order 
to suppress string loops we need to take $N$ large.
For the supergravity description to be valid 't Hooft's 
coupling constant\cite{thooft}, $g_{YM}^2 N$, must be
large. We therefore get that   
the strong ('t Hooft) coupling limit of large $d=4$, $\cn=4$ $SU(N)$ SYM
is described by $adS$ supergravity\cite{malda}!

$\cn=4$ $d=4$ SYM theory is a well-defined unitary finite theory,
whereas supergravity is a non-renormalizable theory. It is best 
to think about it as the low energy effective theory of strings.
Therefore, one should really consider strings on $adS_5 \times S^5$.
In this way we reach the celebrated adS/CFT duality\cite{malda}\footnote{
Many of the elements leading to this conjecture appeared 
in \cite{kleb}. In \cite{KS}, the worldvolume theory 
of the D3 brane was argued to be mapped 
to the singleton of $adS_5$ by the shift transformation.}:

{\it Four dimensional $\cn=4 SU(N)$ SYM is dual to string theory on 
$adS_5 \times S^5$.}

This conjecture was made precise in \cite{GKP,witten2}, where a prescription 
for evaluation of correlation functions was proposed. Subsequently 
a large number of papers appeared, all of them supporting the 
adS/CFT duality.

Let us examine again our result. We obtained that five dimensional
$adS$ gravity is equivalent to $d=4$, $N=4$ SYM theory.
In other words, a gravity theory in $d+1$(=5) dimensions is described 
in terms of a field theory without gravity in $d$(=4) dimensions.
This is just holography\cite{holography1,holography2}! 
One can further show that the boundary theory indeed has 
one degree of freedom per Planck area\cite{wittsuss}.

Similar results hold for other brane configurations as one can always
consider the low energy limit. In the case of conformal worldvolume theories
there is an $adS$ factor on the gravity side. In these
cases the worldvolume theory is valid at all energy scales,
and these considerations provide a weakly coupled gravity description 
of a strongly coupled theory.
In the non-conformal cases the worldvolume SYM theory is 
a theory with a cut-off. As we change the cut-off new degrees
of freedom become relevant
and the description in terms of a SYM theory may not be valid. 
In these cases one finds that as we change the parameters of the 
theory there is always some perturbative description\cite{IMSY,PPol}.
The black $p$-brane solution becomes conformal 
to anti-de Sitter spacetime and the gravity description is in
terms of gauged supergravities which have domain-wall vacua\cite{BST}.

\subsubsection{Low-energy limit of black hole spacetimes}

Let us discuss the low energy limit for black hole configurations.
We will discuss in detail the $5d$ case. The $4d$ case is very 
similar \cite{BaLA}. Rotating black holes have been considered 
in \cite{CvLa}.

Consider the black hole configuration in (\ref{extrBH}).
We go to low energies keeping fixed the masses of stretched strings,
the radius of coordinate which the string is wrapped in 
and the radii of the relative transverse directions in string units,
\be \label{llimit2}
\a' \to 0, \qquad U={r \over \a'}\ \mbox{fixed}, \qquad R_1, 
r_i= {R_i \over \sqrt{\a'}} \ \mbox{fixed}\ \  i=2, \ldots,5
\ee
Notice that $R_1 \gg R_i, i=2, \ldots,5$, as in section \ref{Dbr}. 
Since the horizon is at $r=0$ and $r=U\a' \to 0$
this is at the same time a near-horizon limit. 
Therefore, the resulting configuration has the same number 
of degrees of freedom 
as the original one (since the area of the horizon is 
a measure of the degrees of freedom).

In the limit (\ref{llimit2}) the harmonic functions (\ref{extrHar}) become
\bea
&&H_1 \to {1 \over \a'} {\tilde{Q}_1 \over U^2},\qquad
\tilde{Q}_1={g_s N_1 \over v}, \nonu
&&H_5 \to {1 \over \a'} {\tilde{Q}_5 \over U^2}, \qquad
\tilde{Q}_5=g_s N_5 \nonu
&&K \to 1 + {\tilde{Q}_K \over U^2}, \qquad 
\tilde{Q}_K={g^2_s N_K \over R_1^2 v}
\eea
where $v=r_2 r_3 r_4 r_5$.
Notice that the low-energy limit removes the one from the harmonic
function of the D1 and D5 brane exactly as in section \ref{BTZs}.
Let us define new variables
\bea \label{varia}
&&\r^2 = U^2 + \r_0^2, 
\qquad \f=x_1/R_1, \qquad t_{BTZ}=t {\tilde{Q}_1 \tilde{Q}_5\over R_1^2} \nonu
&&\r_0^2 = \tilde{Q}_K,\qquad l^2={\tilde{Q}_1 \tilde{Q}_5 \over R_1^2}
\eea

The metric (\ref{extrBH}) becomes
\bea \label{lowen}
&&ds^2=\a'{R_1^2 \over \sqrt{\tilde{Q}_1 \tilde{Q}_5}}
[ds_{BTZ}^2 
+{\tilde{Q}_1 \over R_1^2} (dx_2^2 + \cdots +dx_5^2) +
{\tilde{Q}_1  \tilde{Q}_5 \over R_1^2} d \O_3^2] \nonu
&&e^{-2 \f}={\tilde{Q}_5 \over \tilde{Q}_1}
\eea
where $ds_{BTZ}^2$ is the metric (\ref{dsbtz}) with $\r_+=\r_-=\r_0$,
i.e. the metric of the extremal BTZ black hole. The overall factor 
in (\ref{lowen}) originates from the fact that we want to have the 
angle $\f$ with unit radius. We move this overall factor to Newton's
constant by a Weyl rescaling. The three dimensional Netwon's constant
is then equal to (taking into account the dilaton, and arranging such that 
the $3d$ metric is the standard BTZ metric (\ref{dsbtz}))
\be \label{newton}
G_N^{(3)}= {g_s^2 \over 4 R_1 v \sqrt{\tilde{Q}_1 \tilde{Q}_5}}
\ee
Notice that all the factors of $\a'$ have canceled out.
The mass, the angular momentum and the area of the horizon
of the BTZ black hole are equal to 
\be \label{parame}
M= J l, \qquad J = {\r_0^2 \over 4 G_N^{(3)}l}= N_K, \qquad
A=2 \p \r_0 = 2 \p \sqrt{\tilde{Q}_K}
\ee
Therefore,
\be \label{extS1}
S=2 \pi {R_1 v \over g_s^2} \sqrt{\tilde{Q}_1 \tilde{Q}_5 \tilde{Q}_K}
=2 \pi \sqrt{N_1 N_5 N_K}
\ee
as in (\ref{extS}) (as it should since 
we just took the near-horizon limit). Therefore, at low 
energies the physics of extremal black holes is governed by the 
BTZ black hole. 

Let us now move to non-extremal black holes. In this case, the low
energy limit is supplemented by the condition\cite{maldastro},
\be \label{llimit3}
\m_0 = {\m \over \a'}\ \mbox{fixed}
\ee
The non-extremal black hole (\ref{10d}) has an outer horizon 
at $r=\m$ and an inner horizon at $r=0$.
Therefore, the low-energy limit  (\ref{llimit2}), (\ref{llimit3}) 
is a near inner-horizon rather than near outer-horizon limit. 
As a result the entropies do not agree in general. To see this 
observe that the effect of the low energy limit (\ref{llimit2}), 
(\ref{llimit3})
is to remove the one from the harmonic functions $H_1$ and $H_5$ but leave 
$K$ unchanged\cite{maldastro}. Since before we take the 
low energy limit, $H_i(r=\m)=\cosh^2 \a_i$, $i=1,5$
and after the low energy limit $H_i(r=\m)=\sinh^2 \a_i$,
the entropies of the two configurations differ by a factor 
of $\tanh \a_1 \tanh \a_5$. Unless this factor is equal to 
one, the low energy configuration will contain different
number of degrees of freedom. This factor is equal to one in the 
dilute gas approximation
\cite{maldastro1} 
\be
\a_1, \a_5 \gg 1,
\ee
and therefore the entropies agree in this approximation.
Far from extremality the number of degrees of freedom 
changes as we go to low energies.
In all cases the low energy regime is governed by 
the BTZ black hole. This result should be contrasted with the 
result in the previous section. There we also found 
that $4d$ and $5d$ black holes are connected to the BTZ black hole.
All our transformations, however, were isoentropic, and there was 
no limit involved. We only needed that the supergravity approximation
is valid.

Let us finish by presenting a microscopic derivation of the 
Bekenstein-Hawking entropy formula for 
extremal black hole (\ref{extrBH}) using the results of this
section. It has been 
shown by Brown and Henneaux \cite{BrownH}
that the asymptotic symmetry group
of $adS_3$ is generated by two copies of the Virasoro algebra with 
central charge 
\be \label{ccharge}
c={3 l \over 2 G_N^{(3)}}
\ee
This central charge was also derived through the adS/CFT 
correspondence in \cite{HS}.
Therefore, any consistent theory of gravity on $adS_3$ is 
conformal field theory with
central charge equal to (\ref{ccharge}).

The generators of the asymptotic Virasoro are related to the mass and 
angular momentum as 
\bea \label{LL}
&&M={1 \over l}(L_0 + \bar{L}_0), \nonu
&&J=L_0 - \bar{L}_0
\eea
where we have normalized $L_0$, $L_0$ such that they vanish for the 
massless black hole.

In the case of the $5d$ extremal black hole, and after the low-energy limit
is taken, we obtain a geometry of the form $BTZ \times S^3 \times T^4$. 
One may dimensionally reduce over the compact spaces to obtain 
the BTZ black hole and a set of matter fields. The BTZ black hole
is asymptotically $adS_3$ so quantum theory in this space 
is described by a CFT. We can calculate the central charge using
(\ref{varia}), (\ref{newton}). The result is 
\be
c=6 N_1 N_5
\ee
This is the same value as the one we found in section \ref{Dbr}!
In addition, from (\ref{parame}) we obtain $L_0=J=N_K$, $\bar{L}_0=0$. 
Thus, we get the same description as in the D-brane side.
This is the same unitary CFT but we are now at strong coupling.
Therefore, Cardy's formula apply and, (for large black holes,
so $N_K\gg1$) we get correctly (\ref{extS1}).

This counting of states generalizes 
immediately to non-extremal BTZ black holes \cite{BTZstr,BSS}\footnote{
A different counting of the BTZ microstates was presented in 
\cite{carlip}. There it was used the fact that three dimensional gravity 
is topological. The Einstein action can be rewritten as 
a Chern-Simons action\cite{town1,wit}. A Chern-Simons
theory on a manifold with a boundary induces a WZW model in 
the boundary\cite{EMSS}. The degrees of freedom in the boundary 
are would-be gauge degrees of freedom that cannot be 
gauged away because of the boundary. Assuming that the 
horizon is a boundary and imposing certain boundary condition 
one gets that the boundary degrees of freedom can account for the 
black hole entropy\cite{carlip}. A problem with this derivation is that some
of the states counted have negative norms.}. 
(From (\ref{LL}) we get $L_0$, $\bar{L}_0$ in terms of 
$M$ and $J$. We also know $c$ from (\ref{ccharge}). 
Applying Cardy's formula we get the Bekenstein-Hawking 
entropy formula). A crucial point is that in order Cardy's formula to apply
we need the CFT to be unitary. The BTZ black hole, however, induces
a Liouville theory at spatial infinity\cite{carlipL,CHD}.
This means that the effective central charge is equal to 
one\cite{KuSe} instead of $c=2 l/3 G_N^{(3)}$, and one does not 
get correctly the Bekenstein-Hawking entropy formula (see \cite{carlipB}
for further discussion).
We argued that for the case we are discussing we have a unitary CFT 
because of the connection to D-branes. 
We find likely that the CFT corresponding to the BTZ is 
unitary only when the latter is embedded in string theory.

\section*{Acknowledgments}
I would like to thank Jan de Boer for reading the manuscript 
and for discussions and comments.
Research supported by the Netherlands Organization 
for Scientific Research (NWO).


\begin{thebibliography}{999}

\footnotesize
\parskip=0pt plus 1pt

\bibitem{DFR} S. Doplicher, K. Fredenhagen and J. Roberts,
{\it The Quantum Structure of Spacetime at the Planck Scale
and Quantum Fields}, Commun. Math. Phys. {\bf 172} (1995) 187-220.

\bibitem{chris} D. Christodoulou, Phys. Rev. Lett. {\bf 25} (1970) 1596-1597;\\
D. Christodoulou and R. Ruffini, Phys. Rev. {\bf D4} (1971) 3552;\\
R. Penrose and R. Floyd, Nature {\bf 229} (1971) 77; \\
S. Hawking, Phys. Rev. Lett. {\bf 26} (1971) 1344;\\
B. Carter, Nature {\bf 238} (1972) 71.  

\bibitem{Hawk2} J.M. Bardeen, B. Carter and S.W. Hawking,
{\it The four laws of black hole mechanics},
Commun. Math. Phys. {\bf 31} (1973) 161-170.

\bibitem{Beke}J.D. Bekenstein, {\it Black Holes and the Second Law},
Lett. Nuov. Cimento {\bf 4} (1972) 737;
{\it Black Holes and Entropy},
Phys. Rev. {\bf D7} (1973) 2333; {\it Generalized second law of 
thermodynamics in black-hole physics}, Phys. Rev. {\bf D9} (1974) 3292.

\bibitem{Hawk}S.W. Hawking, {\it Black hole explosions?},
Nature {\bf 248} (1974) 30; {\it Particle Creation by Black Holes}
Commun. Math. Phys. {\bf 43} (1975) 199.

\bibitem{HuTo}C.M. Hull and P.K. Townsend, {\it Unity of Superstring 
Dualities} Nucl. Phys. {\bf B438} (1995) 109, hep-th/9410167.

\bibitem{Wit1}
E. Witten, {\it String Theory Dynamics In Various Dimensions}
Nucl. Phys. {\bf B443} (1995) 85, hep-th/9503124.

\bibitem{JPol}J. Polchinski, {\it D-branes and RR-charges}
Phys. Rev. Lett. {\bf 75} (1995) 4724, 
hep-th/9510017. 

\bibitem{BTZ} M. Ba\~{n}ados, C. Teitelboim and J. Zanelli, 
{\it The Black Hole in Three Dimensional Space Time},
\PRL{69}{1992}{1849}, hep-th/9204099. 

\bibitem{BHTZ} M. Ba\~{n}ados, M. Henneaux, C. Teitelboim and J. Zanelli,
{\it Geometry of the $(2+1)$ black hole},
\PRD{48}{1993}{1506}.

\bibitem{Hor1} G. Horowitz, {\it The Dark Side of String Theory: Black Holes 
and Black Strings}, hep-th/9210119.

\bibitem{malda1} J. Maldacena, {\it Black holes in string theory},
hep-th/9607235.

\bibitem{peet} A. Peet, {\it The Bekenstein Formula and String Theory 
(N-brane Theory)}, Class. Quant. Grav. {\bf 15} (1998) 3291-3338,
hep-th/9712253.

\bibitem{TASI} J. Polchinski, {\it TASI Lectures on D-Branes},
hep-th/9611050.

\bibitem{To} P.K. Townsend, 
{\it The eleven-dimensional supermembrane revisited},
Phys. Lett. {\bf B350} (1995) 184-187, hep-th/9501068.

\bibitem{HoWi1}
P. Horava, E. Witten, {\it Heterotic and Type I string dynamics from 
eleven dimensions}
Nucl. Phys. {\bf B460} (1996) 506, hep-th/9510209, and
{\it Eleven-Dimensional Supergravity on a Manifold with Boundary},
Nucl. Phys. {\bf B475} (1996) 94-114, hep-th/9603142.

\bibitem{11sugra}E. Cremmer, B. Julia and J. Scherk, 
{\it Supergravity in eleven dimensions}, Phys. Lett. {\bf B76}
(1978) 409.

\bibitem{Bus1}T. Buscher, {\it A symmetry of the string background 
field equations}, Phys. Lett. {\bf B194} (1987) 59. 

\bibitem{RV} M. Ro\v{c}ek and E. Verlinde, {\it Duality, quotients and 
currents}, Nucl. Phys. {\bf B373} 1992 630-646, hep-th/9110053.

\bibitem{AABL} E. Alvarez, L. Alvarez-Gaume, J.L.F. Barbon and Y. Lozano,
{\it Some global aspects of duality in string theory}, 
Nucl.Phys. {\bf B415} (1994) 71-100, hep-th/9309039.

\bibitem{Bus2}T. Buscher,
{\it Path integral derivation of quantum duality in non-linear sigma models},
Phys. Lett. {\bf B201} (1988) 466.

\bibitem{DRST} J. De Jaegher, J. Raymaekers, A. Sevrin and W. Troost,
{\it Dilaton transformation under abelian and non-abelian T-duality
in the path-integral approach}, hep-th/9812207.

\bibitem{hor1} G. Horowitz and D. Welch, 
{\it Duality Invariance of the Hawking Temperature and Entropy},
Phys. Rev. {\bf D49} (1994) 590,
hep-th/9308077.

\bibitem{Tred} G. Moore, {\it Finite in All Directions}, hep-th/9305139;
C.M. Hull and B. Julia,
{\it Duality and Moduli Spaces for Time-Like Reductions},
Nucl. Phys. {\bf B534} (1998) 250-260, hep-th/9803239;
C.M. Hull, {\it Timelike T-Duality, de Sitter Space, Large $N$ Gauge 
Theories and Topological Field Theory}, J.High Energy Phys. 9807 (1998) 021,
hep-th/9806146.

\bibitem{CLLPST}
E. Cremmer, I.V. Lavrinenko, H. Lu, C.N. Pope, K.S. Stelle and T.A. Tran,
{\it Euclidean-signature Supergravities, Dualities and Instantons},
Nucl. Phys. {\bf B534} (1998) 40-82, hep-th/9803259.

\bibitem{JN}  B. Julia and H. Nicolai,
{\it Null Killing Vector Dimensional Reduction and Galilean Geometrodynamics},
Nucl. Phys. {\bf B439} (1995) 291, hep-th/9412002.

\bibitem{AAB} E. Alvarez, L. Alvarez-Gaume and I. Bakas, 
{\it T-duality and Space-time Supersymmetry}, Nucl. Phys. {\bf B457} 
(1995) 3, hep-th/9507112.

\bibitem{Bak1} 
I. Bakas, {\it Spacetime interpretation of $S$-duality and supersymmetry
violations of $T$-duality},
\PLB{343}{1995}{103}, hep-th/9410104.

\bibitem{BaSf} I. Bakas and K. Sfetsos, 
{\it T-duality and world-sheet supersymmetry},
Phys. Lett. {\bf B349} (1995) 448-457, hep-th/9502065.

\bibitem{GHM} R. Gregory, J.A. Harvey and G. Moore,
{\it Unwinding strings and T-duality of Kaluza-Klein and H-Monopoles},
hep-th/9708086. 

\bibitem{DKL} M.J. Duff, R.R. Khuri, J.X. Lu, {\it String solitons},
Phys. Rept. 259 (1995) 213-326, hep-th/9412184.

\bibitem{stelle} K. S. Stelle, {\it Lectures on Supergravity p-branes},
hep-th/9701088.

\bibitem{Youm} D. Youm, {\it Black Holes and Solitons in String Theory},
hep-th/9710046.

\bibitem{HorStrom} G.T.~Horowitz and A.~Strominger, {\it Black strings
and p-branes}, \NPB{360}{1991}{197}.

\bibitem{BHO} E. Bergshoeff, C.M. Hull and T. Ortin, {\it 
Duality in the Type--II Superstring Effective Action},
Nucl. Phys. {\bf B451} (1995) 547, hep-th/9504081.

\bibitem{coef} R. Myers and M. Perry, {\it Black holes in higher dimensional
spacetimes}, Annals Phys. {\bf 172} (1986) 304. 

\bibitem{dimred} J. Maharana and J. Schwarz, {\it Noncompact
Symmetries in String Theory}, Nucl. Phys. {\bf B390} (1993) 3, hep-th/9207016;
A. Sen, {\it Electric magnetic duality in string theory}, 
Nucl. Phys. {\bf B404} (1993) 109, hep-th/9207053.

\bibitem{witOl} E. Witten and D. Olive, {\it Supersymmetry algebras 
that include topological charges}, Phys. Lett. {\bf 78B} (1978) 97.

\bibitem{DuSt} M.J. Duff and K.S. Stelle, {\it Multimembrane 
solutions of $D=11$ supergravity}, Phys. Lett. {\bf B253} (1991) 113-118.

\bibitem{guven} R. G\"{u}ven, {\it Black $p$-brane solutions of 
$D=11$ supergravity theory}, Phys. Lett. {\bf B276} (1991) 49-55.

\bibitem{PT} G. Papadopoulos and P.K. Townsend, {\it Intersecting $M$-branes},
\PLB{380}{1996}{273}, hep-th/9603087.

\bibitem{Tse1} A.~Tseytlin, {\it Harmonic superpositions of $M$-branes},
\NPB{475}{1996}{149}, hep-th/9604035 and {\it
No-force condition and BPS combinations of $p$-branes in $11$ and $10$
dimensions},
\NPB{487}{1997}{141}, hep-th/9609212.

\bibitem{GKT} J.P.~Gauntlett, D.A.~Kastor and J.~Traschen, {\it
Overlapping Branes in M-Theory},
\NPB{478}{1996}{544}, hep-th/9604179.

\bibitem{ts1} M. Cveti\v{c} and A.A. Tseytlin, 
{\it Non-extreme black holes from non-extreme intersecting M-branes}
Nucl Phys. {\bf B478} (1996) 431, hep-th/9606033.

\bibitem{EREJS} E. Bergshoeff, M. de Roo, E. Eyras, B. Janssen and 
J. P. van der Schaar, {\it Multiple Intersections of D-branes and M-branes},
Nucl. Phys. {\bf B494} (1997) 119-143, hep-th/9612095.

\bibitem{Arefeva} I.Ya. Aref'eva and O.A. Rytchkov, {\it Incidence Matrix 
Description of Intersecting p-brane Solutions}, hep-th/9612236;
I.Ya. Arefeva, K.S. Viswanathan, A.I. Volovich and I.V. Volovich,
{\it Composite p-branes in various dimensions}, 
Nucl. Phys. Proc. Suppl. {\bf 56B} (1997) 52-60, hep-th/9701092;
I.Ya. Aref'eva, M.G. Ivanov and I.V. Volovich, 
{\it Non-extremal Intersecting p-branes in Various Dimensions},
Phys. Lett. {\bf B406} (1997) 44-48, hep-th/9702079.

\bibitem{AEH} R.~Argurio, F.~Englert and L.~Houart, {\it
Intersection Rules for $p$-Branes},
\PLB{398}{1997}{61}, hep-th/9701042.

\bibitem{ohta1} N. Ohta,
{\it Intersection Rules for Non-Extreme $p$-Branes},
Phys. Lett. {\bf B403} (1997) 218-224, hep-th/9702095.

\bibitem{Ber} E. Bergshoeff, M. de Roo, E. Eyras, B. Janssen and
 J.P. van der Schaar, {\it Intersections involving waves and monopoles
 in eleven dimensions}, 
Class.~Quant.~Grav. {\bf 14} (1997) 2757, hep-th/9704120.

\bibitem{ohta2} N. Ohta and J.-G. Zhou,
{\it Towards the Classification of Non-Marginal Bound States of M-Branes
and Their Construction Rules},
Int. J. Mod. Phys. {\bf A13} (1998) 2013-2046,
hep-th/9706153.

\bibitem{ETT} J.D. Edelstein, L. Tataru and R. Tatar,
{\it Rules for localized overlappings and intersections of $p$-branes}, 
High Energy Phys. 9806 (1998) 003, hep-th/9801049.

\bibitem{Gaunt} J.P.~Gauntlett, {\it Intersecting branes}, hep-th/9705011.

\bibitem{khuri} R.R. Khuri, {\it
A Comment on String Solitons},
\PRD{48}{1993}{2947}, hep-th/9305143.

\bibitem{BPS3} H.J. Boonstra, B. Peeters and K. Skenderis, {\it Branes
intersections, anti-de Sitter spacetimes and dual superconformal 
theories}, Nucl. Phys. {\bf B533} (1998) 127-162, hep-th/9803231.

\bibitem{vastro}A. Strominger and C. Vafa, {\it Microscopic 
origin of the Bekenstein-Hawking entropy},
Phys. Lett. {\bf B379} (1996) 99,
hep-th/9601029.

\bibitem{maldaCal}C.G. Callan and J.M. Maldacena, {\it D-brane approach
to black hole quantum mechanics},
Nucl. Phys. {\bf B472} (1996) 591, hep-th/9602043.

\bibitem{hor3} G.T. Horowitz and J. Polchinski, 
{\it A Correspondence Principle for Black Holes and Strings},
Phys. Rev. {\bf D55} (1997) 6189, hep-th/9612146.

\bibitem{KS} K. Sfetsos and K. Skenderis, {\it Microscopic derivation of the 
Bekenstein-Hawking entropy formula for non-extremal black holes},
Nucl. Phys. {\bf B517} (1998) 179-204, hep-th/9711138.

\bibitem{malda} J.M. Maldacena, {\it The Large N Limit of Superconformal 
Field Theories and Supergravity}, Adv. Theor. Math. Phys. {\bf 2}
(1998) 231, hep-th/9711200.

\bibitem{4dBH} R. Kallosh, A. Linde, T. Ortin, A. Peet and A. Van Proeyen,
{\it Supersymmetry as a Cosmic Censor}, Phys. Rev. {\bf D46} (1992) 5278-5302,
hep-th/9205027; M. Cvetic and D. Youm, {\it Dyonic BPS Saturated Black 
Holes of Heterotic String on a Six-Torus}, Phys. Rev. {\bf D53} (1996) 584-588,
hep-th/9507090; M. Cvetic and A.A. Tseytlin, {\it General class of BPS 
saturated dyonic black holes as exact superstring solutions}, 
Phys. Lett. {\bf B366} (1996) 95, hep-th/9510097, and 
{\it Solitonic Strings and BPS Saturated Dyonic Black Holes}, 
Phys. Rev. {\bf D53} (1996) 5619-5633; Erratum-ibid. {\bf D55} (1997) 3907,
hep-th/9512031.

\bibitem{MalStro} J. Maldacena and A. Strominger, {\it Statistical 
Entropy of four-dimensional extremal black holes}, 
Phys. Rev. Lett. {\bf 77} (1996) 428-429, hep-th/9603060.

\bibitem{JKM} C. Johnson, R. Khuri and R. Myers, {\it Entropy of
4$D$ extremal black holes}, Phys. Lett. {\bf B378} 
(1996) 78-86, hep-th/9603061.

\bibitem{rot1}  J.C. Breckenridge, R.C. Myers, A.W. Peet, C. Vafa,
{\it D--branes and Spinning Black Holes}, Phys. Lett. {\bf B391} (1997) 93-98,
hep-th/9602065.

\bibitem{rot2} J. C. Breckenridge, D. A. Lowe, R. C. Myers, A. W. Peet, 
A. Strominger and C. Vafa, {\it Macroscopic and Microscopic Entropy of 
Near-Extremal Spinning Black Holes}, Phys. Lett. {\bf B381} (1996) 423-426,
hep-th/9603078.

\bibitem{rot3} M. Cvetic and D. Youm, {\it Entropy of Non-Extreme 
Charged Rotating Black Holes in String Theory}, Phys. Rev. {\bf D54}
(1996) 2612-2620, hep-th/9603147.

\bibitem{witten1} E. Witten, {\it
Bound States Of Strings And $p$-Branes},
Nucl. Phys. {\bf B460} (1996) 335, 
hep-th/9510135.

\bibitem{sen} A. Sen, {\it A Note on Marginally Stable Bound States 
in Type II String Theory}, Phys. Rev. {\bf D54} (1996) 2964-2967,
hep-th/9510229; and {\it U-duality and Intersecting D-branes},
Phys. Rev. {\bf D53} (1996) 2874-2894, hep-th/9511026.

\bibitem{vaf1} C. Vafa, {\it 
Gas of D-Branes and Hagedorn Density of BPS States},
Nucl. Phys. {\bf B463} (1996) 415-419, hep-th/9511088.

\bibitem{vaf2} M. Bershadsky, V. Sadov and C. Vafa, 
{\it D-Branes and Topological Field Theories}, 
Nucl. Phys. {\bf B463} (1996) 420-434, hep-th/9511222.

\bibitem{vaf3} C. Vafa, {\it Instantons on D-branes},
Nucl. Phys. {\bf B463} (1996) 435-442, hep-th/9512078.

\bibitem{cardy} J.L. Cardy, {\it Operator content of two-dimensional
conformally invariant theories}, Nucl. Phys. {\bf B270} (1986) 186.

\bibitem{SDP}  M. Douglas, J. Polchinski and A. Strominger,
{\it Probing Five-Dimensional Black Holes with D-Branes},
J.High Energy Phys. 9712 (1997) 003, hep-th/9703031.

\bibitem{hyun} S. Hyun, {\it U-duality between Three and Higher 
Dimensional Black Holes}, hep-th/9704005.

\bibitem{BPS1} H.J. Boonstra, B. Peeters and K. Skenderis,
{\it Duality and asymptotic geometries},
\PLB{411}{1997}{59}, hep-th/9706192, and {\it Branes and 
anti-de Sitter spacetimes}, hep-th/9801076.

\bibitem{FerKal} S. Ferrara and R. Kallosh, {\it Supersymmetry and Attractors},
Phys. Rev. {\bf D54} (1996) 1514-1524, hep-th/9602136.

\bibitem{HMS} G. Horowitz, J. Maldacena and A. Strominger,
{\it Nonextremal black hole microstates and U-duality},
Phys. Lett. {\bf B383} (1996) 151-159, hep-th/9603109.

\bibitem{CvHu}  M. Cvetic and C.M. Hull, {\it Black Holes and U-Duality},
Nucl. Phys. {\bf B480} (1996) 296-316, hep-th/9606193.

\bibitem{FeMa} S. Ferrara and J.M. Maldacena, 
{\it Branes, central charges and U-duality invariant BPS conditions},
Class. Quant. Grav. {\bf 15} (1998) 749-758, hep-th/9706097.

\bibitem{hor2}{ G. Horowitz and D. Welch, 
{\it Exact Three Dimensional Black Holes in String Theory},
Phys. Rev. Lett. {\bf 71} (1993) 328, hep-th/9302126.}

\bibitem{berg} E. Bergshoeff and K. Behrndt, 
{\it D-Instantons and asymptotic geometries}, 
Class. Quant. Grav. {\bf 15} (1998) 1801-1813, hep-th/9803090.

\bibitem{Gibb2}  G.W. Gibbons, {\it Wrapping Branes in Space and Time},
hep-th/9803206.

\bibitem{teo} E. Teo, {\it Statistical entropy of charged two-dimensional 
black holes}, Phys. Lett. {\bf B430} (1998) 57-62, hep-th/9803064.

\bibitem{2dBH}  M.D. McGuigan, C.R. Nappi and S.A. Yost,
{\it Charged Black Holes in Two-Dimensional String Theory},
Nucl. Phys. {\it B375} (1992) 421-452, hep-th/9111038; 
G.W. Gibbons and M. Perry, {\it The Physics of 2-d Stringy Spacetimes},
Int.J.Mod.Phys. {\bf D1} (1992) 335-354, hep-th/9204090;
C.R. Nappi and A. Pasquinucci,
{\it Thermodynamics of Two-Dimensional Black-Holes},
Mod. Phys. Lett. {\bf A7} (1992) 3337-3346, gr-qc/9208002.

\bibitem{satoh} Y. Satoh, {\it BTZ black holes and the near-horizon 
geometry of higher-dimensional black holes}, hep-th/9810135.

\bibitem{kaloper}
N. Kaloper, {\it Miens of The Three Dimensional Black Hole}
Phys. Rev. {\bf D48} (1993) 2598, hep-th/9303007.

\bibitem{kumar} A. Ali and A. Kumar,
{\it $O(\tilde d, \tilde d)$ Transformations and 3D Black Hole}
Mod. Phys. Lett. {\bf A8} (1993) 2045-2052.

\bibitem{GPSmonopole}{I. Antoniadis, C. Bachas and A. Sagnotti, {\it
Gauged supergravity vacua in string theory},
Phys. Lett. {\bf B235} (1990) 255;\ 
S.B. Giddings, J. Polchinski and A. Strominger, {\it
Four-dimensional black holes in string theory},
Phys. Rev. {\bf D48} (1993) 5784, hep-th/9305083.}.

\bibitem{BST} H.J. Boonstra, K. Skenderis and P.K. Townsend,
{\it The domain-wall/QFT correspondence}, J.High Energy Phys. 9901 (1999) 003,
hep-th/9807137.

\bibitem{m5} G.W. Gibbons and P.K. Townsend, {\it
Vacuum interpolation in supergravity via super $p$-branes},
Phys.~Rev.~Lett. {\bf 71} (1993) 3754, hep-th/9307049.

\bibitem{thooft} G. 't Hooft, {\it A planar diagram theory 
for strong interactions}, Nucl. Phys. {\bf B72} (1974) 461.

\bibitem{kleb} I.R. Klebanov, {\it World Volume Approach to Absorption 
by Non-dilatonic Branes}, \NPB{496}{1997}{231}, hep-th/9702076; 
S.S. Gubser, I.R. Klebanov, A.A. Tseytlin, 
{\it String Theory and Classical Absorption by Threebranes},
\NPB{499}{1997}{217}, hep-th/9703040;
S.S. Gubser, I.R. Klebanov, {\it Absorption by Branes and Schwinger 
Terms in the World Volume Theory}, \PLB{413}{1997}{41}, hep-th/9708005. 

\bibitem{GKP} S.S. Gubser, I.R. Klebanov and A.M. Polyakov,
{\it Gauge Theory Correlators from Non-Critical String Theory},
Phys. Lett. {\bf B428} (1998) 105, hep-th/9802109.

\bibitem{witten2} E. Witten, {\it Anti-de Sitter space and Holography},
Adv. Theor. Math. Phys. {\bf 2} (1998) 253, hep-th/9802150.

\bibitem{holography1} G. 't Hooft, {\it Dimensional reduction in Quantum
gravity}, Class. Quant. Grav. {\bf 11} (1994) 621, gr-qc/9310006.

\bibitem{holography2} L. Susskind, {\it The World as a Hologram},
J. Math. Phys. {\bf 36} (1995) 6377, hep-th/9409089.

\bibitem{wittsuss} L. Susskind and E. Witten,
{\it The Holographic Bound in Anti-de Sitter Space}, hep-th/9805114.

\bibitem{IMSY} N. Itzhaki, J. Maldacena, J. Sonnenschein and S. Yankielowicz,
{\it Supergravity and The Large N Limit of Theories With Sixteen Supercharges},
Phys. Rev. {\bf D58} (1998) 046004,
hep-th/9802042.

\bibitem{PPol} A.W. Peet, J. Polchinski, {\it UV/IR Relations in AdS Dynamics},
hep-th/9809022.

\bibitem{BaLA} V. Balasubramanian and F. Larsen,
{\it Near Horizon Geometry and Black Holes in Four Dimensions}
\NPB{528}{1998}{229}, hep-th/9802198.

\bibitem{CvLa} M. Cvetic and F. Larsen, {\it Near Horizon Geometry of 
Rotating Black Holes in Five Dimensions}, Nucl. Phys. {\bf B531} 
(1998) 239-255 and {\it Microstates of Four-Dimensional Rotating 
Black Holes from Near-Horizon Geometry}, hep-th/9805146.

\bibitem{maldastro} J. Maldacena and A. Strominger, {\it $AdS_3$ black 
holes and a stringy exclusion principle}, hep-th/9804085. 

\bibitem{maldastro1} J. Maldacena, A. Strominger,
{\it Black Hole Greybody Factors and D-Brane Spectroscopy},
Phys. Rev. {\bf D55} (1997) 861-870, hep-th/9609026.

\bibitem{BrownH} J.D. Brown and M. Henneaux, 
{\it Central charges in the canonical realization 
of asymptotic symmetries: An example from three-dimmensional 
gravity}, Commun. Math. Phys. {\bf 104} (1986) 207.

\bibitem{HS} M. Henningson and K. Skenderis,
{\it The Holographic Weyl anomaly},  J.High Energy Phys. 9807 (1998) 023,
hep-th/9806087; {\it Holography and the Weyl anomaly},
hep-th/9812032.

\bibitem{BTZstr} A. Strominger, 
{\it Black Hole Entropy from Near-Horizon Microstates},
J.High Energy Phys. 9802 (1998) 009, hep-th/9712251.

\bibitem{BSS}  D. Birmingham, I. Sachs and S. Sen,
{\it  Entropy of Three-Dimensional Black Holes in String Theory},
Phys. Lett. {\bf B424} (1998) 275-280, hep-th/9801019.

\bibitem{carlip} S. Carlip, 
{\it The Statistical Mechanics of the (2+1)-Dimensional Black Hole},
Phys. Rev. {\bf D51} (1995) 632, gr-qc/9409052 and  
{\it The Statistical Mechanics of the Three-Dimensional Euclidean Black Hole},
{\bf D55} (1997) 878, gr-qc/9606043.

\bibitem{carlipL} S. Carlip, {\it Inducing Liouville theory 
from topologically massive gravity}, Nucl. Phys. {\bf B362} (1991) 111-124.

\bibitem{CHD} O. Coussaert, M. Henneaux and P. van Driel,
{\it The asymptotic dynamics of three-dimensional Einstein gravity
with negative cosmological constant}, Class. Quant. Grav. {\bf 12}
(1995), 2961-2966.

\bibitem{town1} A. Ach\'{u}carro and P. K. Townsend, 
{\it A Chern-Simons action for three-dimensional anti-de Sitter
supergravity theories},
Phys. Lett. {\bf B180} (1986) 89. 

\bibitem{wit} E. Witten, 
{\it $(2+1)$-dimensional gravity as an exactly soluble system},
Nucl. Phys. {\bf B311} (1988) 46. 

\bibitem{EMSS} G. Moore and  N. Seiberg, {\it Taming the conformal zoo},
Phys. Lett. {\bf B220} (1989) 422; 
S. Elitzur, G. Moore, A. Schwimmer and
Nathan Seiberg, {\it Remarks on the canonical quantization of the
Chern-Simons-Witten theory}, Nucl. Phys. {\bf B326} (1989) 108.

\bibitem{carlipB} S. Carlip, {\it What We Don't Know about BTZ Black 
Hole Entropy}, Class. Quant. Grav. {\bf 15} (1998) 3609-3625, hep-th/9806026.

\bibitem{KuSe} D. Kutasov and N. Seiberg, {\it Number of degrees of freedom,
density of states and tachyons in string theory and CFT},
Nucl. Phys. {\bf B358} (1991) 600-618.











\end{thebibliography}
\end{document}